\pdfoutput=1
\documentclass[12pt]{article}

\font\grande=cmr9.5 scaled \magstep4
\font\medio=cmr9.5 scaled \magstep2
\outer\def\beginsection#1\par{\medbreak\bigskip
      \message{#1}\leftline{\bf#1}\nobreak\medskip
\vskip-\parskip
      \noindent}
\usepackage{graphicx} 
\usepackage{mathrsfs}
\begin{document}

\bibliographystyle{unsrt}

\titlepage

\vspace{1cm}
\begin{center}
{\grande Effective anisotropic stresses of the relic gravitons}\\
\vspace{1cm}
Massimo Giovannini \footnote{e-mail address: massimo.giovannini@cern.ch}\\
\vspace{1cm}
{{\sl Department of Physics, CERN, 1211 Geneva 23, Switzerland }}\\
\vspace{0.5cm}
{{\sl INFN, Section of Milan-Bicocca, 20126 Milan, Italy}}
\vspace*{1cm}
\end{center}
\vskip 0.3cm
\centerline{\medio  Abstract}
\vskip 0.1cm
The effective anisotropic stresses induced by the scalar modes of the geometry 
depend on the coordinate system so that the comparison of the competing results 
is ultimately determined by the evolution of the pivotal variables in each particular gauge. 
After arguing that the only reasonable physical coordinate systems for this problem are 
the ones where the gauge freedom is completely fixed (like the longitudinal  and the uniform 
curvature gauges), we propose a novel gauge-invariant strategy for the comparison of 
gauge-dependent results.  Instead of employing the pivotal variables of a given coordinate system, 
the effective anisotropic stress is solely expressed in terms of the gravitating normal modes of the plasma 
and in terms of their conformal time derivatives. The new approach is explicitly gauge-invariant
and when the wavelengths of the normal modes are either shorter or larger than the sound horizon, 
the physical limits of the anisotropic stresses are determined without relying on the specific details of the background evolution. 
The relevance of the proposed strategy is discussed in the general situation where the scalar anisotropic stress and the non-adiabatic 
pressure fluctuations are simultaneously present. We finally argue that  the  anisotropic stress can be most efficiently obtained from the second-order effective action 
of the curvature inhomogeneities.  
\noindent
\vspace{5mm}
\vfill
\newpage

\renewcommand{\theequation}{1.\arabic{equation}}
\setcounter{equation}{0}
\section{Introduction}
\label{sec1}
The effective energy densities and the pressures of the relic gravitons 
are neither unique nor gauge-invariant.  This inevitable feature is ultimately caused by the equivalence 
principle that forbids the localization of the energy-momentum tensor 
of the gravitational field \cite{ONE}. The {\em effective anisotropic stresses} 
induced by the scalar inhomogeneities of the geometry also affect
the evolution of the relic gravitons and they are customarily assessed always by 
using the Landau-Lifshitz approach \cite{TWO} with the proviso that
besides the second-order tensor modes (leading to the energy-momentum 
pseudo-tensor) also the second-order scalar modes should be consistently 
taken into account. Within the concordance paradigm the spectral energy density 
of the relic gravitons scales linearly with the amplitude of the tensor 
power spectrum (i.e. ${\mathcal A}_{T}$) while the correction due to the 
effective anisotropic stresses coming from the scalar modes is quadratic in the amplitude 
of the scalar power spectra ${\mathcal A}_{{\mathcal R}}$.  The corrections coming from the scalar 
anisotropic stresses are therefore smaller than the leading-order results  by a factor where 
${\mathcal A}_{{\mathcal R}}/r_{T}$ where $r_{T}$ denotes the tensor to scalar ratio
(see, for instance, \cite{TWOa} for a recent review). There are two related aspects that make this problem 
often confusing. First the Landau-Lifshitz approach is not unique;
second the effective anisotropic stresses are, by construction, gauge dependent.
 In this investigation we shall address both issues with the aim of 
proposing a novel gauge-invariant approach to the analysis 
of the effective anisotropic stresses induced by the scalar modes of the 
geometry.

It is actually well known that the Landau-Lifshitz strategy \cite{TWO} is not unique: the Brill-Hartle 
averaging \cite{THREE},  the Isaacson approach \cite{FOUR,FIVE} and the Ford-Parker 
proposal \cite{SIX} are the main suggestions put forward through the years 
for a proper definition of the energy-momentum  pseudo-tensor of the gravitational field. 
As recently argued these approaches are not always equivalent:
if applied in a cosmological context different proposals lead to sharply different forms of the energy density 
and of the pressure of the relic gravitons \cite{SEVEN}. If the frequencies of the gravitons 
are larger than the rate of variation of the background the different pseudo-tensors 
lead to coincident results but the conclusions are sharply different in the opposite limit.
 
While some other suggestions have been presented through the years, they can all be 
related, either directly or indirectly, to the original ideas mentioned in the previous paragraph. 
 So for instance the proposal of Ref. \cite{EIGHT} coincides with the Landau-Lifshitz  approach while the 
 results of Refs. \cite{EIGHTA,EIGHTB} follow from the strategies of Refs. \cite{THREE,FOUR,SIX}. 
 The suggestion of Refs. \cite{EIGHTC}  coincide with the approach of the effective action \cite{SIX} (see also \cite{FIVE}).  
 The authors of Ref. \cite{NINE} claimed a result with all the necessary properties of a true energy-momentum tensor of the 
gravitational field itself (i.e. symmetry, uniqueness, gauge-invariance and covariant conservation). 
While this result has been subsequently challenged by Refs. \cite{TEN,ELEVEN},
 the geometrical object most closely related to the suggestion \cite{NINE} is the Landau-Lifshitz pseudo-tensor \cite{TWO}.
The ambiguity of the competing definitions may be solved by imposing a number of physical requirements 
(e.g. the positivity of the energy density both inside and outside the Hubble radius) \cite{SEVEN}. 
These criteria pin down the Ford-Parker proposal \cite{SIX} where the energy-momentum pseudo-tensor 
follows from the variation of the effective action of the relic gravitons with respect to the background metric.

The possibility of higher-order processes makes the problem more acute and, in  a sense, even less 
gauge-invariant. For instance the long-wavelength gravitons induce curvature inhomogeneities both 
during inflation and in the subsequent radiation-dominated phase \cite{TWELVE}. 
Similarly curvature inhomogeneities may cause higher-order corrections to the stochastic backgrounds of 
relic gravitons and this second effect involves an effective anisotropic stress \cite{THIRTEEN}. The
 gauge-dependence of the effective anisotropic stresses has been originally 
suggested in Ref. \cite{SIXTEEN}. Even if the 
description of the longitudinal gauge is  considered more computable 
and hence more reliable (see e. g. \cite{SEVENTEEN}), there are no reasons why 
this should be the case so that the scheme of Ref. \cite{SIXTEEN} has been subsequently replicated
with different and sometimes contradictory conclusions \cite{EIGHTEEN,NINETEEN,TWENTY}.
Reference \cite{EIGHTEEN} attributes the difference of the results 
to the evolutionary features of each gauge. Reference \cite{NINETEEN} suggests
that the effective anisotropic stress is gauge-independent but the authors also imply, in their conclusions,
that  the observational sensitivities for the tensor perturbations induced from the effective anisotropic stress 
will be different from those for conventional gravitational waves. In this sense the observation of this tensor perturbation 
might require a discussion about the suitable gauge for the observation because of its gauge dependence. 
This last statement is at odds with the claimed gauge-independence. Finally Ref.  \cite{TWENTY} overlaps significantly 
with previous works and, by admission of the authors, 
it just revisits the gauge dependence of gravitational waves generated at second order from scalar perturbations.
This analysis suggests that the various backgrounds affect the gauge-invariance of the final results and claims that the 
obtained conclusions are not really gauge-independent and, to some extent, even background dependent. 

We propose here a method that is simultaneously gauge-invariant and background independent. 
The idea is to obtain the effective anisotropic stresses in a particular gauge and then to express the 
obtained result solely in terms of  ${\mathcal R}$ and ${\mathcal R}^{\prime}$
that will denote throughout  the curvature inhomogeneities (defined on comoving orthogonal hypersurfaces) 
and their corresponding (conformal) time derivatives. Since the new variables coincide with the gravitating normal modes of the system
their evolution is the same in any gauge. Therefore, within the present approach, the effective anisotropic 
stresses in different gauges will depend {\em on the same set of pivotal variables 
obeying the same master equation}: unlike the strategies pursued so far the comparison between the gauge-dependent 
results will therefore be immediate. In Refs. \cite{TWENTYONEa} and \cite{TWENTYONEb} the main 
aspects of this approach have been outlined in the simplest possible situation, namely 
the one where the non-adiabatic pressure fluctuations are absent and the total anisotropic stress 
vanishes. 

To avoid potential confusions we stress that 
three different quantities shall be repeatedly mentioned hereunder namely: 
\begin{itemize}
\item{} the anisotropic stress induced by free-streaming particles ($\Pi_{t}$ in what follows) and affecting the evolution 
of the {\em scalar} modes of the geometry;
\item{} the effective anisotropic stress induced by the (second-order) scalar inhomogeneities and affecting the evolution 
of the {\em tensor} modes;
\item{} the non-adiabatic pressure fluctuations ($\delta p_{nad}$ in what follows) depend on the composition 
of the plasma and it vanishes in the case of a single fluid. 
\end{itemize}
The anisotropic stress caused by the free-streaming particles (for short the {\em scalar anisotropic stress}) in the concordance paradigm is mainly due to
neutrinos and it affects the initial conditions of the Einstein-Boltzmann hierarchy 
necessary for the calculation of the temperature and polarization 
anisotropies of the Cosmic Microwave Background  \cite{TWENTYONEc}.
The neutrinos free-stream after electron-positron annihilation and their anisotropic stress also affects directly
the relic graviton background by suppressing its spectral energy density \cite{TWENTYFOURa,TWENTYFOURb}.
The second-order scalar modes of the geometry induce instead an effective anisotropic stress which is the one 
considered more directly here.  The non-adiabatic pressure fluctuations vanish in the case of the concordance 
paradigm but may contribute to more general scenarios both at early and at late times. 

Also in the presence of $\delta p_{nad}$ and $\Pi_{t}$ the effective anisotropic can be solely expressed in 
terms ${\mathcal R}$ and ${\mathcal R}^{\prime}$ but the non-adiabatic pressure fluctuations and the scalar anisotropic stress 
will however introduce a source term in the evolution equation for the gravitating normal modes of the system. 
 An important technical advantage of the present approach concerns the approximate 
solutions of the evolution that can be analyzed in a background-independent manner. Indeed the single master equation obeyed by
${\mathcal R}$ and ${\mathcal R}^{\prime}$  can be analyzed within the Wentzel-Kramers-Brillouin 
 (WKB) approximation.   
   
All in all the layout of this investigation is the following. In section \ref{sec2} we shall present the gauge-invariant 
  evolution of the gravitating normal modes of the plasma when the non-adiabatic pressure 
  fluctuations and the total anisotropic stress are present. The general properties of the 
  effective anisotropic stresses will also be outlined with particular attention to the 
  coordinate systems where the gauge freedom is completely fixed. 
  In sections \ref{sec3} and \ref{sec4}  the main idea will be illustrated
 by explicitly deriving the expressions of the effective 
  anisotropic stress in terms of the gauge-invariant normal modes. 
  In particular the longitudinal gauge will be discussed  in section \ref{sec3} while section \ref{sec4} will be instead 
  focussed on the coordinate system where the spatial curvature is uniform.  It will be shown that different gauge-invariant descriptions (like the one following 
from the density contrast on uniform curvature hypersurfaces) cannot be traded for 
the one based on ${\mathcal R}$ and ${\mathcal R}^{\prime}$. In section \ref{sec5} the gauge-dependent results will be 
  compared in gauge-invariant terms with particular attention 
  to the limits of the effective anisotropic stress for typical 
  wavelengths larger or shorter than the sound horizon. 
In section \ref{sec6} the spectral energy density of relic gravitons will be 
  computed in the case of the concordance paradigm and for a radiation-dominated plasma. In section \ref{sec7} we shall clarify how the effective anisotropic stress 
  could be derived from the second-order action of the scalar modes in full analogy with the procedure 
leading to the effective energy density of the relic gravitons. Section \ref{sec8} contains the concluding considerations.

\renewcommand{\theequation}{2.\arabic{equation}}
\setcounter{equation}{0}
\section{General gauge-invariant evolution}
\label{sec2}
\subsection{Gravitating normal modes}
In a conformally flat and homogeneous background geometry the fluctuations of a gravitating, irrotational and relativistic fluid admit a normal mode
that shall be conventionally denoted hereunder by ${\mathcal R}$. This quantity has been originally discussed by Lukash \cite{TWENTYTWO}  even prior to the actual formulation of the conventional inflationary paradigm and in the context of the pioneering analyses of the relativistic theory of large-scale inhomogeneities \cite{TWENTYTWOa,TWENTYTWOb}.  There are different situations where the evolution of ${\mathcal R}$ can be 
studied. In the simplest case the non-adiabatic pressure fluctuations and the scalar anisotropic stress are absent. 
The evolution of ${\mathcal R}$ obeys then the following decoupled equation:
\begin{equation}
{\mathcal R}'' + 2 \frac{z_{t}'}{z_{t}} {\mathcal R}' - c_{\mathrm{st}}^2 \nabla^2 {\mathcal R}  =0,
\label{one1}
\end{equation}
where the prime denotes a derivation with respect to the conformal time coordinate $\tau$ which is related to the cosmic 
time as $a(\tau) d\tau = dt$; in Eq. (\ref{one1})  $c_{st}^2$ and $z_{t}$ are defined as:
\begin{equation}
 c_{\mathrm{st}}^2 = \frac{p_{t}'}{\rho_{t}'}, \qquad z_{t} = \frac{a^2 \sqrt{p_{t} + \rho_{t}}}{{\mathcal H} c_{\mathrm{st}}}.
\label{one2}
\end{equation}
From Eq. (\ref{one2})  $z_{t}$ and $c_{s}^2$ depend on the the total energy density $\rho_{t}$ and  on the total pressure $p_{t}$; 
moreover, using standard notations, ${\mathcal H} = a^{\prime}/a = a H$ and $H$ denotes the Hubble 
expansion rate. In the absence of further sources ${\mathcal H}$, $p_{t}$ and $\rho_{t}$ 
will obey the conventional Friedmann-Lema\^itre equations 
\begin{equation}
3 {\mathcal H}^2 = \ell_{P}^2 \,a^2 \, \rho_{t}, \qquad 2({\mathcal H}^2 - {\mathcal H}^{\prime} )= \ell_{P}^2 \, a^2\, (\rho_{t} + p_{t}),
\label{one2a}
\end{equation}
where $\ell_{P} = \sqrt{ 8 \pi G}$.  The variable ${\mathcal R}$ deduced in Ref. \cite{TWENTYTWO} and obeying Eq. (\ref{one1}) coincides with the curvature perturbation on comoving orthogonal hypersurfaces and it is invariant under infinitesimal coordinate transformations as required in the context of the Bardeen formalism \cite{TWENTYTWOc}. Subsequent analyses  \cite{TWENTYTWOd,TWENTYTWOe} followed the same logic of \cite{TWENTYTWO} but in the case of scalar field matter.  All the normal modes identified in Refs. \cite{TWENTYTWO,TWENTYTWOd,TWENTYTWOe}  can be related to the (rescaled) curvature perturbations on comoving orthogonal hypersurfaces \cite{TWENTYTWOf}. 

\subsection{Non-adiabatic pressure fluctuations}
Equation (\ref{one1}) is obtained by assuming the absence of the non-adiabatic pressure fluctuations 
and the absence of any source of anisotropic stress due to free-streaming particles. We are now going 
to relax both hypotheses. The non-adiabatic pressure 
fluctuations \cite{TWENTYTWOg,TWENTYTWOh,TWENTYTWOi}  arise for several reasons even 
if in the context of  the concordance paradigm they are bound to vanish. In general terms the pressure fluctuations may not be only caused by the inhomogeneities of the energy density of the plasma, as it happens in the concordance paradigm, 
 so that the pressure perturbation shall be written as the sum of two 
different contributions:
\begin{equation}
\delta_{s} p_{t} = \, c_{st}^2 \, \delta_{s} \rho + \delta p_{nad}, \qquad \delta p_{nad}(\vec{x},\tau) = \sum_{a\,b}\frac{\partial p_{\mathrm{t}}}{\partial \varsigma_{a\,b}} \delta \varsigma_{a\,b}(\vec{x},\tau).
\label{one3}
\end{equation}
While the first term of Eq. (\ref{one3}) accounts for the fluctuations of the 
pressure coming from the inhomogeneity of the energy density, 
the explicit expression of $\delta p_{and}$ depends on the composition 
of the plasma and it vanishes in the case of a single fluid. In Eq. (\ref{one3}) $\varsigma_{a\, b}$ denotes the specific entropy, i.e.
the ratio between the entropy density and the concentration of the given species; the indices $a$ and $b$ denote instead 
the various species of the plasma. The entropy fluctuation is therefore defined as the relative fluctuation
of the specific entropy for a given pair of species in the plasma:
\begin{equation}
{\mathcal S}_{a\, b}(\vec{x},\tau) = \frac{\delta \varsigma_{a\, b}(\vec{x},\tau)}{\varsigma_{a\,b}}= \frac{\delta_{b}}{w_{b} + 1} - 
\frac{\delta_{a}}{w_{a}+1}, \qquad {\mathcal S}_{a\, b} = - {\mathcal S}_{b\, a}
\label{one3a}
\end{equation}
While the first expression in Eq. (\ref{one3a}) follows from the definition, the second equality holds when the different species 
with the density contrasts $\delta_{a}= \delta\rho_{a}/\rho_{a}$ and $\delta_{b}=\delta\rho_{b}/\rho_{b}$ 
are characterized by the constant barotropic indices $w_{a}$ and $w_{b}$.  For a collection of fluids with different equations 
of state and different sound speeds the explicit form of $\delta p_{and}$ is
\begin{equation}
\delta p_{nad}(\vec{x},\tau)  = \frac{1}{6 {\mathcal H} \rho_{\mathrm{t}}'} \sum_{a\,b} \rho_{a}' \rho_{b}' 
(c_{s\, a}^2 - c_{s\, b}^2) {\mathcal S}_{a\, b}(\vec{x},\tau),\qquad 
{\mathcal S}_{a\, b}(\vec{x},\tau) = \frac{\delta \varsigma_{a\, b}(\vec{x},\tau)}{\varsigma_{a\,b}},
\label{one4}
\end{equation}
where, as in Eq. (\ref{one3a}) the indices $a$ and $b$ are not tensor indices but denote 
two generic species of the pre-equality plasma; $c_{s\, a}^2= p_{a}^{\prime}/\rho_{a}^{\prime}$ and $c_{s\, b}^2= p_{b}^{\prime}/\rho_{b}^{\prime}$ are the corresponding sound speeds. In the case of a fluid made of two different components 
(e.g. radiation and matter) the corresponding total energy density is $\rho_{t} = (\rho_{M} + \rho_{R})$ 
with $\rho_{M}' = - 3 {\mathcal H} \rho_{M}$ and $\rho_{R}^{\prime} = - 4 {\mathcal H} \rho_{R}$. 
From Eq. (\ref{one4}) the explicit expression of $\delta p_{nad}$ will be given by:
\begin{equation}
\delta p_{nad} = \frac{4}{3}\,\frac{\rho_{M} \rho_{R}}{ 4 \rho_{R} + 3 \rho_{M}} {\mathcal S}_{*},\qquad {\mathcal S}_{*}= {\mathcal S}_{M\, R} = - {\mathcal S}_{R\, M}.
\label{one4a}
\end{equation}
Note that ${\mathcal S}_{*}$ can also be expressed as 
the fractional variation of the specific entropy $\varsigma = T_{R}^3/n_{M}$ where 
$T_{R}$ is the temperature of the radiation background and $n_{m}$ is concentration 
of matter species; indeed we have $\delta\varsigma/\varsigma= (3 \delta_{R}/4 - \delta_{M})$
where $\delta_{R} = \delta \rho_{R}/\rho_{R}$ and $\delta_{M}=\delta\rho_{M}/\rho_{M}$; exactly the same 
result follows from Eq. (\ref{one3a}) 
From the expressions of $\rho_{M}$ and $\rho_{R}$ in terms of the scale factor Eq. (\ref{one4a}) 
becomes
\begin{equation}
\delta p_{nad} = \rho_{M} \, c_{st}^2 \, {\mathcal S}_{*}, \qquad  c_{st}^2 = \frac{p_{t}^{\prime}}{\rho_{t}^{\prime}} = \frac{4}{3[ (a/a_{*})+ 4]},
\label{one4b}
\end{equation}
where it is understood that the plasma is dominated by radiation for $a> a_{*}$ and by matter for $a< a_{*}$.
In the conventional terminology \cite{TWENTYTWOg,TWENTYTWOh,TWENTYTWOi} 
Eqs. (\ref{one4a}) and (\ref{one4b}) describe either the CDM-radiation mode (if $\rho_{M} = \rho_{cdm}$) or the 
baryon--radiation mode (provided $\rho_{M} = \rho_{baryon}$). 
In the concordance paradigm, when the dark energy does not fluctuate, there are, overall five different sets of Cauchy data: one adiabatic  and four non-adiabatic \cite{TWENTYTWOg,TWENTYTWOh,TWENTYTWOi} initial conditions\footnote{On top of the  CDM-radiation mode and of the baryon-entropy mode the remaining two non-adiabatic modes are the neutrino entropy mode and the neutrino isocurvature velocity mode. The considerations discussed hereunder are not bound to the case of the illustrative examples of 
Eqs. (\ref{one4a}) and (\ref{one4b}) but apply for all the non-adiabatic solutions. }.

\subsection{Quasi-normal modes}
Equation (\ref{one1}) also  neglects the scalar anisotropic stress due to free-streaming particles (e.g. neutrinos
in the case of the concordance paradigm).
The total anisotropic stress associated with the scalar modes will therefore be expressed in one of the 
following equivalent ways:
\begin{equation}
\partial_{i} \,\partial_{j} \, \Pi_{t}^{i \, j} = \nabla^2 \Pi_{t}, \qquad \Pi_{t} = (\rho_{t} + p_{t}) \, \sigma_{t}.
\label{one5}
\end{equation}
If the plasma contains a total anisotropic stress and non-adiabatic pressure fluctuations Eq. (\ref{one1}) 
gets modified by inheriting a source term ${\mathcal S}_{{\mathcal R}}(\vec{x},\tau)$:
\begin{equation}
{\mathcal R}'' + 2 \frac{z_{t}'}{z_{t}} {\mathcal R}' - c_{\mathrm{st}}^2 \nabla^2 {\mathcal R}  = {\mathcal S}_{{\mathcal R}}(\vec{x},\tau).
\label{one6}
\end{equation}
If $\delta p_{nad} \neq 0$ and $\Pi_{t} \neq 0$ the source term ${\mathcal S}_{{\mathcal R}}$ can be written in the following matter:
\begin{eqnarray}
{\mathcal S}_{{\mathcal R}}(\vec{x},\tau) &=&  \Sigma_{\mathcal R}' + 2 \frac{z_{t}'}{z_{t}} \Sigma_{\mathcal R}+ \frac{ 3 a^{4}}{z_{t}^2 } \Pi_{t},
\label{one7}\\
\Sigma_{{\mathcal R}}(\vec{x},\tau) &=& - \frac{{\mathcal H}}{p_{t} + \rho_{t}} \delta p_{nad} + \frac{{\mathcal H}}{p_{t} + \rho_{t}} \Pi_{t}.
\label{one8}
\end{eqnarray}
 In Fourier space Eq. (\ref{one6}) becomes therefore:
\begin{equation}
\biggl({\mathcal R}_{\vec{k}}^{\prime} - \Sigma_{\vec{k}}\biggr)^{\prime} + 
2 \frac{z_{t}^{\prime}}{z_{t}} \biggl({\mathcal R}_{\vec{k}}^{\prime} - \Sigma_{\vec{k}}\biggr)  +k^2 \, c_{\mathrm{st}}^2 {\mathcal R}_{\vec{k}} = \frac{3 \, a^4}{z_{t}^2} \Pi_{\vec{k}} 
\label{one15}
\end{equation}
where $\Pi_{\vec{k}}$ is the Fourier transform of $\Pi_{t}$;
$\Sigma_{\vec{k}}$ and $\Gamma_{\vec{k}}$ are instead the Fourier transforms of $\Sigma_{{\mathcal R}}$ and 
$\delta p_{nad}$:
\begin{equation}
\Sigma_{\vec{k}} = \frac{{\mathcal H}}{p_{t} + \rho_{t}} \biggl(\Pi_{\vec{k}} - \Gamma_{\vec{k}}\biggr)
\label{one16}
\end{equation}
While non-adiabatic pressure fluctuations and the anisotropic stress have been given in general terms, in the context 
of the concordance paradigm $\delta p_{nad} =0$ and the total anisotropic stress is only due to neutrinos:
\begin{equation}
\Pi_{t} = (p_{\nu} +\rho_{\nu}) \sigma_{\nu}, \qquad \sigma_{\nu} = \frac{{\mathcal F}_{\nu}}{2}.
\label{one17}
\end{equation}
In Eq. (\ref{one17}) ${\mathcal F}_{\nu\,2}$ is the quadrupole of the neutrino phase-space distribution. The lower moments of ${\mathcal F}_{\nu \ell}$ 
(i.e. with $\ell =0,\, 1$)  are related with the density contrast and with the peculiar velocity of the neutrinos while for $\ell \geq 3$ 
the evolution of ${\mathcal F}_{\nu\ell}$ is given by:
\begin{equation}
{\mathcal F}_{\nu\ell}'= \frac{k}{2\ell + 1} \biggl[ \ell {\mathcal F}_{\nu(\ell -1)} - (\ell + 1) {\mathcal F}_{\nu(\ell+1)}\biggr],\qquad \ell\geq 3.
\label{one18}
\end{equation}
The evolution of the anisotropic stress of the neutrinos can be obtained by cutting the Boltzmann hierarchy of Eq. (\ref{one18})
and by requiring, for instance, ${\mathcal F}_{\nu\,3} =0$ (but, according to Eq. (\ref{one18}), ${\mathcal F}_{\nu\, 3}^{\prime} \neq 0$).
To get a decoupled equation we have to pay the price of higher derivatives of $\sigma_{\nu}$ and the result is:
\begin{eqnarray}
&& \sigma_{\nu}''' + \frac{8}{5} {\mathcal H}^2 R_{\nu} \Omega_{R} \sigma_{\nu}' + \frac{6}{7} k^2 \sigma_{\nu}' - \frac{32}{5}  {\mathcal H}^3 R_{\nu} \Omega_{R} \sigma_{\nu} 
\nonumber\\
&&= \frac{8}{15 c_{\mathrm{st}}^2} \biggl( {\mathcal H} - \frac{{\mathcal H}'}{{\mathcal H}}\biggr) \biggl(\frac{{\mathcal H}'}{{\mathcal H}} - 2 {\mathcal H}\biggr) ({\mathcal R}' - \Sigma_{{\mathcal R}} ) + \frac{8}{15} \biggl(\frac{{\mathcal H}'}{{\mathcal H}} - {\mathcal H}\biggr) \, k^2 {\mathcal R},
\label{one19}
\end{eqnarray}
where $\Omega_{R} = \rho_{R}/\rho_{t}$ is the critical fraction of radiation; as anticipated $R_{\nu}$ and $R_{\gamma}$ count the fraction of neutrinos and photons in the radiation plasma. 

\subsection{Effective anisotropic stresses of the relic gravitons}
The effective anisotropic stress of the relic gravitons follows 
by perturbing the Einstein equations as 
\begin{equation}
\delta_{t}^{(1)} {\mathcal G}_{i}^{\,\,\,j} + \delta_{s}^{(2)} {\mathcal G}_{i}^{\,\,\,j} = \ell_{P}^2\, \delta_{s}^{(2)} T_{i}^{\,\,j}.
\label{onea1}
\end{equation}
 Equation (\ref{onea1}) follows directly from the Landau-Lifshitz strategy \cite{TWO} and few notational comments are in order:
\begin{itemize}
\item{} ${\mathcal G}_{\mu}^{\nu}$ will denote throughout the Einstein tensor while$T_{\mu}^{\nu}$ is a generic energy-momentum tensor of the matter sources; for the present discussion we shall be mostly concerned with the 
case of hydrodynamical matter where $T_{\mu}^{\nu} = (\rho_{t} + p_{t}) u_{\mu} u^{\nu} - p_{t} \delta_{\mu}^{\nu}$;
\item{} in Eq. (\ref{onea1}) $\delta_{t}^{(1)}$ denotes the first-order 
tensor fluctuation while $\delta_{s}^{(2)}$ denotes the second-order scalar of the corresponding 
quantities;
\item{}  at the left-hand-side of  Eq. (\ref{onea1}) there is a further contribution 
coming from the second-order tensor fluctuations ${\mathcal G}_{\mu}^{\,\,\nu}$, i.e. $\delta_{t}^{(2)} {\mathcal G}_{\mu}^{\,\,\nu}$: 
note in fact that $- \delta^{(2)}_{t} {\mathcal G}_{\mu\nu}/\ell_{P}^2$ is nothing but the Landau-Lifshitz pseudo-tensor \cite{TWO} (see also \cite{SEVEN} for different ways 
of assigning the energy density and pressure of the relic gravitons).
\end{itemize}
While $\delta_{t}^{(1)} {\mathcal G}_{i}^{\,\,\,j}$ is gauge-invariant to first-order, the second-order contributions are 
both gauge-dependent. With this proviso, since the explicit expression of $\delta_{t}^{(1)} {\mathcal G}_{i}^{\,\,\,j} $ is:
\begin{equation}
\delta_{t}^{(1)} {\mathcal G}_{i}^{\,\,\,j} = - \frac{1}{2 a^2} \biggl(  h_{i}^{\,\,j\,\prime\prime} + 2 {\mathcal H} \, h_{i}^{\,\,j\,\prime}  - \nabla^2 h_{i}^{\,\,j} \biggr), 
\label{onea2}
\end{equation}
Eq. (\ref{onea1}) can also be expressed as 
\begin{equation}
h_{i}^{\,\,j\,\prime\prime} + 2 {\mathcal H} \, h_{i}^{\,\,j\,\prime}  - \nabla^2 h_{i}^{\,\,j}  = - 2 \ell_{P}^2 a^2 \Pi_{i}^{(X)\,\,j},
\label{onea3}
\end{equation}
where $\Pi_{i}^{(X)\,\,j}$ now defines the effective anisotropic stress  determined from the scalar 
fluctuations of the geometry and computed in the gauge $X$:
\begin{equation}
\Pi_{i}^{(X)\,\,j} = \delta_{s}^{(2)} T_{i}^{(X)\,\,j} - \frac{1}{\ell_{P}^2} \delta_{s}^{(2)} {\mathcal G}_{i}^{(X)\,\,j}.
\label{onea4}
\end{equation}
Equation (\ref{onea3}) is ambiguous: while at the left-hand side the tensor part is formally gauge-invariant, 
the effective anisotropic stress is instead gauge-dependent so different anisotropic stresses, computed in diverse 
coordinate systems will determine different tensor amplitudes which should be instead coordinate-independent. 
This is, in a nutshell, one of the motivations
of the present analysis: to avoid manifest contradictions  it is important to find a gauge-invariant method to compare various gauge-dependent 
results. 

The effective anisotropic stress $\Pi_{i}^{(X)\,\,j}$ appearing in Eqs. (\ref{onea3}) and (\ref{onea4}) 
{\em is determined up to total spatial derivatives involving quadratic combinations of the pivotal variables of a given gauge}. This property is a direct consequence 
of the Landau-Lifshitz approach leading to Eqs. (\ref{onea1}) and (\ref{onea2}). Therefore, given 
a quadratic combination of two first-order fluctuations (e.g. $Q$ and $P$) in a specific gauge, 
the identity 
\begin{equation}
\partial_{i} Q\, \partial^{j} P = - Q \partial_{i} \, \partial^{j} P + \partial_{i} (  Q \partial^{j} P),
\label{onea5}
\end{equation}
can always be used with the aim of neglecting the second term at the right-hand side. This is possible since the effective anisotropic stress must be always projected along the two tensor polarizations and, in this process, the total derivative 
of Eq. (\ref{onea5}) carries a comoving three-momentum $q^{i}$ which is orthogonal to both tensor polarizations.
To clarify this point we recall that the Fourier transforms of $h_{i}^{\,\,j}(\vec{x},\tau)$ and $\Pi_{i}^{(X)\,\,j}(\vec{x},\tau)$ are defined as:
\begin{equation}
h_{i}^{\,\,j}(\vec{q},\tau) = \frac{1}{(2\pi)^{3/2}} \int d^3 x \, h_{i}^{\,\,j}(\vec{x},\tau), \qquad \Pi_{i}^{(X)\,\,j}(\vec{q},\tau) = 
 \frac{1}{(2\pi)^{3/2}} \int d^3 x \, \Pi_{i}^{(X)\,\,j}(\vec{x},\tau).
\label{onea6}
\end{equation}
If the Fourier amplitude is expanded in the basis of the tensor polarizations we obtain:
 \begin{equation}
h_{i}^{\,\,\, j} (\vec{q}, \tau) =\sum_{\lambda= \oplus,\, \otimes} \, e^{(\lambda)\,\,\,j}_{i}(\hat{q}) \, \, h_{\lambda}(\vec{q},\tau), \qquad \Pi_{i}^{(X)\,\,\,j}(\vec{q},\tau) = \sum_{\lambda= \oplus,\, \otimes}  \, e^{(\lambda)\,\,\,j}_{i}(\hat{q}) \, \Pi_{\lambda}^{(X)}(\vec{q},\tau).
\label{onea8} 
\end{equation}
In Eq. (\ref{onea8})  $e_{i\,j}^{(\oplus)}(\hat{q})$ and $e_{i\,j}^{(\otimes)}(\hat{q})$ are given by:
\begin{equation}
 e_{i\,j}^{(\oplus)}(\hat{q}) = \hat{m}_{i} \,\hat{m}_{j} - \hat{n}_{i} \,\hat{n}_{j}, \qquad e_{i\,j}^{(\otimes)}(\hat{q}) = \hat{m}_{i} \,\hat{n}_{j} + \hat{n}_{i} \,\hat{m}_{j},
 \label{onea7}
\end{equation} 
where $\hat{m}$, $\hat{n}$ and $\hat{q}$ are three mutually orthogonal unit vectors. Using Eq. (\ref{onea8}) 
Eq. (\ref{onea3}) becomes, in Fourier space,
\begin{equation}
h_{\lambda}^{\prime\prime} + 2 {\mathcal H} h_{\lambda}^{\prime} + q^2 h_{\lambda} = - 2 \,\ell_{P}^2 \,a^2(\tau) \, \Pi_{\lambda}^{(X)}.
\label{onea9}
\end{equation}
Equations (\ref{onea8}) and (\ref{onea9}) imply then that total spatial derivatives [like the second 
term at the right hand side of Eq. (\ref{onea5})] will not contribute to $\Pi_{\lambda}^{(X)}(\vec{q},\tau)$
since they will always be orthogonal both to $ e_{i\,j}^{(\oplus)}(\hat{q})$ and to $ e_{i\,j}^{(\otimes)}(\hat{q})$ [i.e.
$\hat{q}^{i}  e_{i\,j}^{(\oplus)}(\hat{q}) =  \hat{q}^{i} e_{i\,j}^{(\otimes)}(\hat{q}) =0$].

\subsection{Gauges for the effective anisotropic stresses and their drawbacks}

Since the effective anisotropic stress of the relic gravitons must be evaluated in a particular gauge 
the potential presence of spurious gauge modes should be avoided. These unwanted modes arise 
when the gauge freedom is not completely removed and they mix with the evolution of the physical modes by often 
making unphysical the obtained expressions of the effective anisotropic stresses. This drawback is already present to first-order (see e.g. \cite{TWENTYFIVEa}) but it becomes even more acute when dealing with quadratic combinations of the 
perturbations variables in a given gauge, as it happens for the explicit evaluation of the effective anisotropic stresses.
For this purpose we recall that the scalar fluctuations of the $(3+1)$-dimensional metric are parametrized by four 
independent functions which can be eventually reduced by specifying (either completely or partially) the coordinate system: 
\begin{equation}
 \delta_{\mathrm{s}}^{(1)} g_{00}(\vec{x},\tau) = 2 a^2 \phi, \qquad \delta_{\mathrm{s}}^{(1)} g_{ij}(\vec{x},\tau) = 2 a^2(\psi \delta_{ij} - E_{i\,j}), \qquad 
 \delta^{(1)}_{\mathrm{s}} g_{0i}(\vec{x},\tau) = - a^2 V_{i},
\label{onea10} 
\end{equation}
where, in the scalar case, $V_{i} = \partial_{i} B$ and $E_{ij} = \partial_{i} \partial_{j}E$.
For infinitesimal coordinate shifts  of the type:
\begin{equation}
\tau \to \widetilde{\,\tau\,} = \tau + \epsilon_{0},\qquad  {x}^{i} \to \widetilde{\,x\,}^{i} = x^{i} + \partial^{i}\epsilon,
\label{onea10aa}
\end{equation}
 the functions $\phi(\vec{x},\tau)$, $B(\vec{x},\tau)$, 
$\psi(\vec{x},\tau)$ and $E(\vec{x},\tau)$ introduced in Eq. (\ref{onea10}) transform as:
\begin{eqnarray}
&& \phi \to \widetilde{\,\,\phi\,\,} = \phi - {\mathcal H} \epsilon_0 - \epsilon_{0}' ,\qquad \psi \to \widetilde{\,\,\psi\,\,} = \psi + {\mathcal H} \epsilon_{0},
\label{onea11}\\
&& B \to \widetilde{\,\,B\,\,} = B +\epsilon_{0} - \epsilon',\qquad E \to \widetilde{\,\,E\,\,} = E - \epsilon.
\label{onea12}
\end{eqnarray}
Two commonly employed coordinate systems where the gauge freedom  is completely fixed are the 
conformally Newtonian (or longitudinal) gauge where $E=0$ and $B=0$ and the off-diagonal 
(or uniform curvature) gauge where $E=0$ and $\psi=0$. In fact if we start from the situation 
where $E \neq 0$ and $B\neq 0$ the longitudinal condition $\widetilde{E}= \widetilde{B}=0$ can be always recovered by setting 
\begin{equation}
\epsilon(\vec{x},\tau) = E(\vec{x},\tau), \qquad \epsilon_{0} = E^{\prime}(\vec{x},\tau) - B(\vec{x},\tau).
\label{onea12aa}
\end{equation}
Similarly if we start from the situation 
where $E \neq 0$ and $\psi \neq 0$ the off-diagonal coordinate system $\widetilde{E} = \widetilde{\psi} =0$ follows by setting 
\begin{equation}
\epsilon(\vec{x},\tau) = E(\vec{x},\tau), \qquad \epsilon_{0}(\vec{x},\tau) = - \frac{\psi(\vec{x},\tau)}{{\mathcal H}}.
\label{onea12bb}
\end{equation}
In the case of Eqs. (\ref{onea12aa}) and (\ref{onea12bb}) the coordinate system is completely fixed. Conversely 
there are gauges where the gauge freedom can only be fixed up to arbitrary (space-dependent) constants.
For instance the synchronous coordinate system  is defined  by $\phi =0$ and $B=0$ and if we start 
from a physical situation where the synchronous condition is not verified (i.e. $\phi \neq 0$ and $B\neq 0$)
the condition $\widetilde{\phi} = \widetilde{B} = 0$ can only be satisfied up to two arbitrary constants. 
Indeed from Eqs. (\ref{onea11}) and (\ref{onea12}) we see that the condition $\widetilde{\phi} = \widetilde{B} = 0$
is recovered provided:
\begin{eqnarray}
\epsilon_{0}(\vec{x},\tau) &=& \frac{C_{1}(\vec{x})}{a(\tau)} + \frac{1}{a(\tau)} \int_{0}^{\tau} \phi(\vec{x},\tau_{1}) \, d\tau_{1},
\nonumber\\
\epsilon(\vec{x},\tau) &=& C_{2}(\vec{x}) + C_{1}(\vec{x}) \int_{0}^{\tau} \frac{d\tau_{1}}{a(\tau_{1})} +
\int_{0}^{\tau} B(\vec{x},\tau_{1}) \, d\tau_{1} + \int_{0}^{\tau} \frac{d\tau_{1}}{a(\tau_{1})} \int_{0}^{\tau_{1}} \phi(\vec{x},\tau_{2}) \, d\tau_{2},
\label{onea12cc}
\end{eqnarray}
From Eq. (\ref{onea12cc}) it is apparent that  the synchronous gauge condition is not 
completely fixed unless $C_{1}(\vec{x})$ and $C_{2}(\vec{x}) $ are specified. This overall ambiguity 
causes the presence of  spurious gauge modes \cite{TWENTYFIVEa}. This problem is 
potentially even more acute in the case of the effective anisotropic stresses and, for this reason,
 the illustrative considerations of the following two sections shall mainly involve those coordinate systems where 
the gauge freedom is completely fixed.

When the coordinate system is completely fixed the  individual linear order variables used
in one gauge cannot immediately compared to the ones of another gauge  and this 
is especially true in the case of the effective anisotropic stresses containing 
quadratic combinations of the metric inhomogeneities. The variables $\phi$ and $\psi$ 
in the $L$-gauge (or $\phi$ and $B$ in the $U$-gauge) are {\em not gauge-invariant} 
since they take a different form when the coordinate system changes. Conversely ${\mathcal R}$ 
and ${\mathcal R}^{\prime}$ obey the same equation in any coordinate system. This is, in a nutshell, 
the advantage of working directly with the gravitating normal modes of the plasma\footnote{Note that 
$\phi$ in the $U$-gauge and in the $L$-gauge is different insofar as it obeys different 
equations.}.

\renewcommand{\theequation}{3.\arabic{equation}}
\setcounter{equation}{0}
\section{The longitudinal gauge picture}
\label{sec3}
The standard approach to the analysis of the effective anisotropic stresses of the relic gravitons relies on gauge-dependent 
treatments. By this we mean that not only the anisotropic stress is computed in a specific gauge but that 
also the evolution of the various variables is followed {\em in that specific coordinate system}. The idea pursued here
is different: instead of studying and evolving the effective anisotropic stresses in terms of the pivotal variables 
of a specific coordinate system we express the pivotal variables of that gauge in terms of the curvature perturbations 
and of their first-order derivatives with respect to the conformal time coordinate.  Among the possible gauges 
where the effective anisotropic stresses could be computed, only the ones where the gauge freedom is 
completely fixed guarantee the absence of spurious gauge modes. For this reason in the present section we shall first examine 
the  longitudinal picture while in the following section the uniform curvature 
gauge will be more specifically analyzed.  Recalling Eqs. (\ref{onea10}), (\ref{onea11})--(\ref{onea12}) and (\ref{onea12aa}), 
in the longitudinal gauge the metric fluctuations are expressed as: 
\begin{equation}
\delta_{s}^{(1)} g_{00}(\vec{x},\tau) = 2 a^2 \, \phi, \qquad  \delta_{s}^{(1)} g_{ij}(\vec{x},\tau) = 2 a^2 \,\psi \delta_{ij}. 
\label{two1}
\end{equation}
 In the standard approach the effective anisotropic stress is computed in terms 
of $\phi$ and $\psi$ so that effective anisotropic stress depends on the evolutionary features 
of the longitudinal gauge. Since our aim is to compare the effective anisotropic stresses in different gauges the idea 
is to trade the pivotal variables of a given gauge for the curvature inhomogeneities.  So, for instance, the relation between the curvature perturbations on comoving orthogonal hypersurfaces and the longitudinal degrees of freedom (\ref{two1}) in Fourier space is given by:
\begin{eqnarray}
&& {\mathcal R}_{\vec{k}} = - \psi_{\vec{k}} - \frac{{\mathcal H}  ( {\mathcal H} \phi_{\vec{k}} + \psi_{\vec{k}}^{\prime})}{{\mathcal H}^2 - {\mathcal H}^{\prime}},
\label{two2}\\
&& {\mathcal R}_{\vec{k}}^{\prime} = \Sigma_{\vec{k}} + \frac{2\, a^2\, k^2\, \psi_{\vec{k}}}{ \ell_{P}^2 {\mathcal H} z_{t}^2}.
\label{two3}
\end{eqnarray}
The accuracy of Eqs. (\ref{two2}) and (\ref{two3}) can be immediately verified by checking that they lead to the equation 
of the the quasinormal modes already discussed in Eq. (\ref{one6}). Since Eq. (\ref{one6}) is gauge-invariant it can be 
derived in any gauge and, in particular, in the gauge (\ref{two1}). Let us therefore derive once both sides of Eq. (\ref{two2}) with respect to the conformal time coordinate $\tau$; if we then use, in the obtained 
expression, Eqs. (\ref{two1}) and (\ref{two2}) we arrive at the following expression 
\begin{equation}
{\mathcal R}_{\vec{k}}^{\prime\prime} + 2 \frac{z_{t}'}{z_{t}} {\mathcal R}_{\vec{k}}^{\prime} = \Sigma_{\vec{k}}' + 2 \frac{z_{t}'}{z_{t}} \Sigma_{\vec{k}} 
- \frac{a^2 k^2 ({\mathcal H}^2 - {\mathcal H}')}{ 4 \pi G {\mathcal H} z_{t}^2} {\mathcal R} + 
\frac{k^2 a^2 {\mathcal H}}{4 \pi G {\mathcal H} z_{t}^2} (\psi_{\vec{k}} - \phi_{\vec{k}}),
\label{two4}
\end{equation}
where the only dependence on the longitudinal fluctuations of the metric 
is in the last term. We then recall that, in the gauge (\ref{two1}), the scalar anisotropic stress discussed in Eq. (\ref{one5}) 
accounts for the mismatch between the two longitudinal fluctuations of the metric. In Fourier space we then have 
\begin{equation}
k^2 (\phi_{\vec{k}} - \psi_{\vec{k}}) = \Delta_{\vec{k}}, \qquad \Delta_{\vec{k}} = - \frac{3}{2} \ell_{P}^2\, a^2 \Pi_{\vec{k}}.
\label{two5}
\end{equation}
Inserting Eq. (\ref{two5}) into Eq. (\ref{two4}) the obtained result coincides, as expected, with Eq. (\ref{one15}).
We stress that in Eq. (\ref{two5}) we  introduced, for the sake of conciseness, $\Delta_{\vec{k}}$ which is only a convenient auxiliary quantity.

\subsection{The effective anisotropic stress in terms of $\phi$ and $\psi$} 

In the $L$-gauge of Eq. (\ref{two1}) the effective anisotropic stress given in Eq. (\ref{onea4}) follows from the standard Landau-Lifshitz approach and it is formally expressed as:
\begin{equation}
\Pi_{i}^{\,\,(L) \,j} = \delta_{s}^{(2)} \, T_{i}^{\,\,(L)\, j} - \frac{1}{\ell_{P}^2} \delta_{s}^{(2)} {\mathcal G}_{i}^{\,\,(L)\, j}. 
\label{two6}
\end{equation}
The second-order fluctuation of the sources appearing in the first term at the right-hand side of Eq. (\ref{two6}) is easily computed by recalling that, in the $L$-gauge, 
\begin{equation}
\delta_{s} u_{i} = \frac{2}{a \ell_{P}^2 (p_{t} + \rho_{t})} \partial_{i}\biggl( \psi + {\mathcal H} \phi\biggr). 
\label{two7}
\end{equation}
Equation (\ref{two7}) follows from the first-order  fluctuation of the Einstein equations with mixed indices 
in the longitudinal gauge.  Neglecting the trace we therefore have that $\delta_{s}^{(2)} \, T_{i}^{\,\,(L)\, j} $ is given by:
\begin{eqnarray}
\delta_{s}^{(2)} \, T_{i}^{\,\,(L)\, j}  = (\rho_{t} + p_{t}) \delta_{s} u_{i} \, \delta_{s} u^{j} 
= - \frac{2}{a^2 \,\ell_{P}^2 \,({\mathcal H}^2 - {\mathcal H}^{\prime})} \partial_{i}\biggl({\mathcal H} \phi + \psi^{\prime} \biggr) \partial^{j} \biggl({\mathcal H} \phi + \psi^{\prime} \biggr).
\label{two8}
\end{eqnarray}
Similarly, always neglecting the trace, $\delta_{s}^{(2)} \, {\mathcal G}_{i}^{\,\,(L)\, j} $ is given by:
\begin{eqnarray}
\delta_{s}^{(2)} \, {\mathcal G}_{i}^{\,\,(L)\, j} = \frac{1}{a^2} \biggl[ \partial_{i} \phi \partial^{j} \phi - \partial_{i} \psi \partial^{j}  \psi 
- 2 \psi \partial_{i} \partial^{j} \biggl(\phi - \psi\biggr) + \partial_{i} \phi \partial^{j} \psi + \partial_{i} \psi \partial^{j} \phi\biggr].
\label{two9}
\end{eqnarray} 
Putting together the obtained results the effective anisotropic stress of Eq. (\ref{two6}) becomes 
\begin{eqnarray}
\Pi_{i}^{\,\,(L) \,j}(\vec{x},\tau) &=& - \frac{1}{a^2 \ell_{P}^2} \biggl[ \partial_{i} \phi \partial^{j} \phi -  \partial_{i} \psi \partial^{j} \psi +
\partial_{i} \phi \partial^{j} \psi +  \partial_{i} \psi \partial^{j} \phi 
\nonumber\\
&-& 2 \psi \partial_{i} \partial^{j}(\phi - \psi) + \frac{2}{({\mathcal H}^2 - {\mathcal H}^{\prime})} \partial_{i}\biggl({\mathcal H} \phi + \psi^{\prime} \biggr) \partial^{j} \biggl({\mathcal H} \phi + \psi^{\prime} \biggr)\biggr].
\label{two10}
\end{eqnarray}
The result of Eq. (\ref{two10}) follows by recalling that the enthalpy density of the background (i.e. $p_{t} + \rho_{t}$) 
can always eliminated thanks to Eq. (\ref{one2a}).  Equation (\ref{two10}) is then further simplified thanks to Eq. (\ref{onea5}): since the effective anisotropic stress will be eventually 
projected along the two tensor polarizations the total spatial derivatives do not contribute 
to the final expression. In particular in the $L$-gauge Eq. (\ref{onea5}) implies:
\begin{equation}
\partial_{i} \phi \partial^{j} \psi = - \phi \, \partial_{i} \partial^{j} \psi + \partial_{i} \biggl(\phi \, \partial^{j} \psi \biggr).
\label{two11}
\end{equation}
Thanks to Eq. (\ref{two11}) the effective anisotropic stress of Eq. (\ref{two10}) becomes:
\begin{eqnarray}
\Pi_{i}^{\,\,(L) \,j}(\vec{x},\tau) &=&  \frac{1}{a^2 \ell_{P}^2} \biggl[  \phi \partial_{i} \partial^{j} \phi -  \partial_{i} \psi \partial^{j} \psi +
\phi \partial_{i} \partial^{j} \psi +  \psi \partial_{i} \partial^{j} \phi 
\nonumber\\
&+& 2 \psi \,\partial_{i} \partial^{j}(\phi - \psi) + \frac{2}{({\mathcal H}^2 - {\mathcal H}^{\prime})} \biggl({\mathcal H} \phi + \psi^{\prime} \biggr)\partial_{i}\, \partial^{j} \biggl({\mathcal H} \phi + \psi^{\prime} \biggr)\biggr].
\label{two12}
\end{eqnarray}
Finally, in Fourier space Eq. (\ref{two12}) reads:
\begin{eqnarray}
\Pi_{i}^{\,\,(L) \,j}(\vec{q},\tau) &=& \frac{1}{(2\pi)^{3/2}} \int e^{i \, \vec{q}\cdot\vec{x}} \, \Pi_{i}^{\,\,(L) \,j}(\vec{x},\tau)
\nonumber\\ 
&=& - \frac{1}{ (2 \pi)^{3/2}\,a^2 \, \ell_{P}^2 } \int d^{3} k \, k_{i} \, k^{j} \,\biggl[ \phi_{\vec{q} - \vec{k}} \phi_{\vec{k}} - \psi_{\vec{q} - \vec{k}} \psi_{\vec{k}} + \phi_{\vec{q} - \vec{k}} \psi_{\vec{k}} + 
\psi_{\vec{q} - \vec{k}} \phi_{\vec{k}} 
\nonumber\\
&+& 2 \psi_{\vec{q} - \vec{k}} ( \phi_{\vec{k}} - \psi_{\vec{k}} )
+ \frac{2}{{\mathcal H}^2 - {\mathcal H}^{\prime}} \biggl({\mathcal H} \phi_{\vec{q}- \vec{k}} + \psi_{\vec{q}- \vec{k}}^{\prime} \biggr) \biggl({\mathcal H} \phi_{\vec{k}} + \psi_{\vec{k}}^{\prime} \biggr)\biggr].
\label{two13}
\end{eqnarray}

\subsection{The effective anisotropic stress in terms of ${\mathcal R}$ and ${\mathcal R}^{\prime}$}

Equation (\ref{two13}) can be directly studied in terms of $\phi_{\vec{k}}$ and $\psi_{\vec{k}}$ which are the pivotal 
variables of the $L$-gauge.
This is what has been done in previous studies but this approach is not ideal for a sound physical 
comparison of the results obtained in different gauges. If this strategy is strictly followed the anisotropic stresses derived in different 
gauges can only be compared at the very end and also for specific classes of background evolutions. The aim of this analysis 
is opposite: we would like to compare the different results before specifying the evolution of the background. The idea is therefore 
to use Eqs. (\ref{two2})--(\ref{two3}) by trading $\phi_{\vec{k}}$ and $\psi_{\vec{k}}$ for ${\mathcal R}_{\vec{k}}$ and ${\mathcal R}_{\vec{k}}$: 
\begin{equation}
\phi_{\vec{k}} = \psi_{\vec{k}} + \Delta_{\vec{k}},\qquad 
\psi_{\vec{k}} = \frac{{\mathcal H}^2 - {\mathcal H}^{\prime}}{{\mathcal H} \, k^2 \, c_{st}^2} \biggl({\mathcal R}_{\vec{k}}^{\prime} - \Sigma_{\vec{k}}\biggr)
\label{two14}
\end{equation} 
Using the above expression the effective anisotropic stress can be expressed as:
\begin{eqnarray}
\Pi_{i\,j}^{(L)}(\vec{q},\tau) &=& - \frac{2 ( {\mathcal H}^2 - {\mathcal H}^{\prime})}{ (2\pi)^{3/2}\, \ell_{P}^2\, a^2\, {\mathcal H}^2} 
\int d^{3} k\,\, k_{i} k_{j} \biggl[ {\mathcal R}_{\vec{k}} {\mathcal R}_{\vec{q} -\vec{k}}  +  \frac{{\mathcal H}^2}{2 ({\mathcal H}^2 - {\mathcal H}^{\prime})} \Delta_{\vec{k}} \, \Delta_{\vec{q} - \vec{k}}
\nonumber\\
&+&\frac{ ( {\mathcal H}^2 - {\mathcal H}^{\prime}) ( 2 {\mathcal H}^2 - {\mathcal H}^{\prime})}{ k^2 \, {\mathcal H}^2 \, |\vec{q} - \vec{k}|^2 \, c_{st}^4} \, ({\mathcal R}_{\vec{k}}^{\prime} - \Sigma_{\vec{k}}) ({\mathcal R}_{\vec{q} - \vec{k}}^{\prime} - \Sigma_{\vec{q} - \vec{k}}) 
\nonumber\\
&+& \frac{3 {\mathcal H}}{2\,k^2 c_{st}^2} \Delta_{\vec{q} - \vec{k}}({\mathcal R}_{\vec{k}}^{\prime} - \Sigma_{\vec{k}})
+ \frac{3 {\mathcal H}}{2\,|\vec{q} - \vec{k}|^2 c_{st}^2} \Delta_{\vec{k}} ({\mathcal R}_{\vec{q} - \vec{k}}^{\prime} - \Sigma_{\vec{q} - \vec{k}})
\nonumber\\
&+& \frac{({\mathcal H}^2 - {\mathcal H}^{\prime})}{{\mathcal H} \,|\vec{q} - \vec{k}|^2 \, c_{st}^2} {\mathcal R}_{\vec{k}} 
({\mathcal R}_{\vec{q} - \vec{k}}^{\prime} - \Sigma_{\vec{q} - \vec{k}})
+ \frac{({\mathcal H}^2 - {\mathcal H}^{\prime})}{{\mathcal H} \,k^2 \, c_{st}^2} {\mathcal R}_{\vec{q} -\vec{k}} 
({\mathcal R}_{\vec{k}}^{\prime} - \Sigma_{\vec{k}})
\biggr].
\label{two15}
\end{eqnarray}
The advantage of Eq. (\ref{two15}) in comparison with Eq. (\ref{two13}) is evident: while $\phi_{\vec{k}}$ and $\psi_{\vec{k}}$ 
obey the equations that are specific to the $L$-gauge, ${\mathcal R}_{\vec{k}}$ and ${\mathcal R}_{\vec{k}}^{\prime}$ 
obey instead Eq. (\ref{one15}) that has the same form in any coordinate system (i.e. its is gauge-invariant). 
In previous studies (see for instance Ref. \cite{TWENTY}) the curvature inhomogeneities have been used to normalise the results 
obtained in different gauges. This procedure is, strictly speaking, background-dependent insofar as ${\mathcal R}_{\vec{k}}$ 
is taken to be strictly constant. Some of these approaches (like the one of Ref. \cite{TWENTY})
are only consistent  in the case where the curvature inhomogeneities are time-independent and cannot be used
in the general situation where, on the contrary, Eq. (\ref{two15}) applies without approximations.

We conclude this part of the discussion by remarking that the gauge-invariant evolution
of the neutrino anisotropic stress of Eq. (\ref{one19}) can also be obtained directly in the 
$L$-gauge of Eq. (\ref{two1}). In other words, starting form the lowest multiples 
of the neutrino hierarchy we can easily deduce Eq. (\ref{one19}) directly in the longitudinal gauge. In short the derivation is the following.
Recalling that the  lowest multipoles of the neutrino hierarchy read, in the longitudinal gauge,
\begin{eqnarray}
&& \delta_{\vec{k}}^{\prime} = - \frac{4}{3} \theta_{\vec{k}} + 4 \psi_{\vec{k}}^{\prime},
\label{two16}\\
&&\theta_{\vec{k}}^{\prime} =  \frac{k^2}{4} \delta_{\vec{k}} - k^2 \sigma_{\nu} + k^2 \phi_{\vec{k}},
\label{two17}\\
&& \sigma_{\nu}^{\prime} = \frac{4}{15} \theta_{\vec{k}} - \frac{3}{10} k {\mathcal F}_{\nu 3},
\label{two18}
\end{eqnarray}
where $\delta_{\vec{k}}$ is the neutrino density contrast and $\theta_{\vec{k}}$ is the three-divergence 
of the corresponding peculiar velocity. 
If we take the conformal time derivative of both sides of Eq. (\ref{two16}); we thus obtain 
\begin{equation}
\sigma_{\nu}^{\prime\prime} = \frac{k^2}{15} \delta_{\vec{k}} + \frac{4}{15} k^2 \phi_{\vec{k}} - \frac{11}{21}  k^2 \sigma_{\nu},
\label{two19}
\end{equation}
where the neutrino hierarchy has been truncated, for illustration, to the octupole (notice, however, that ${\mathcal F}_{\nu\, 3}' \neq 0$).
From Eq. (\ref{two19}) it also follows that:
\begin{equation}
\sigma_{\nu}^{\prime\prime\prime} + \frac{6}{7} k^2 \sigma_{\nu}' = \frac{4 k^2}{15} (\phi_{\vec{k}} -\psi_{\vec{k}})^{\prime} + \frac{8}{15} k^2 \psi_{\vec{k}}^{\prime}.
\label{two20}
\end{equation}
In Eq. (\ref{two20}) the term $k^2(\phi_{\vec{k}} - \psi_{\vec{k}})^{\prime}$ can be replaced by taking the derivative of both sides 
of Eq. (\ref{two5}); the other term appearing at  the right hand side of Eq. (\ref{two20}) is instead replaced by taking the 
derivative of Eq. (\ref{two3}) and by inserting, in the obtained expression, the decoupled equation for ${\mathcal R}_{\vec{k}}$, i.e. Eq. (\ref{one6}).
The result in terms of $k^2 \psi_{\vec{k}}^{\prime}$ becomes:
\begin{equation}
k^2 \psi_{\vec{k}}^{\prime} = \frac{1}{c_{\mathrm{st}}^2} \biggl({\mathcal H} - \frac{{\mathcal H}'}{{\mathcal H}}\biggl) (  \frac{{\mathcal H}'}{{\mathcal H}} - 2{\mathcal H} ) ({\mathcal R}_{\vec{k}}' - \Sigma_{\vec{k}})  
+ 6 {\mathcal H}^3 \Omega_{R} R_{\nu}  \sigma_{\nu}  - \biggl({\mathcal H} - \frac{{\mathcal H}'}{{\mathcal H}}\biggr) k^2 {\mathcal R}_{\vec{k}}.
\label{two21}
\end{equation}
If Eq. (\ref{two21}) is now plugged into Eq. (\ref{two20}) we obtain the 
equation already reported in Eqs. (\ref{one19}) provided the term $k^2(\phi_{\vec{k}} - \psi_{\vec{k}})^{\prime}$ is eliminated by means of the derivative of Eq. (\ref{two3}).

\subsection{Complementary gauge-invariant descriptions}

Equation (\ref{two15}) demonstrates that the effective anisotropic stress can be expressed directly 
in terms of ${\mathcal R}_{\vec{k}}$ and ${\mathcal R}_{\vec{k}}^{\prime}$ not only asymptotically (i.e. when 
${\mathcal R}_{\vec{k}}^{\prime} \to 0$) but in general terms. There could be 
some suggesting that ${\mathcal R}$ should also be traded for another popular gauge-invariant 
variable conventionally denoted by $\zeta$. The gauge-invariant relation between the two variables 
is:
\begin{equation}
 \zeta - {\mathcal R} = \frac{2 \nabla^2 \psi}{ 3 \ell_{P}^2 a^2 ( p_{t} + \rho_{t})} =  \frac{\Sigma_{\mathcal R} - {\mathcal R}'}{3 {\mathcal H} c_{\mathrm{st}}^2}.
\label{two22}
\end{equation} 
The first equality in Eq. (\ref{two22}) holds in the $L$-gauge while the second relation is gauge-invariant. 
Equation (\ref{two22}) shows that if we would trade ${\mathcal R}$ for $\zeta$ we should also generate 
new terms proportional to ${\mathcal R}^{\prime}$. In the $L$-gauge the explicit definition of $\zeta$ is given by 
\begin{equation}
\zeta = -   \psi - {\mathcal H} \frac{\delta \rho_{\mathrm{t}} }{\rho_{\mathrm{\mathrm{t}}}'}.
\label{two23}
\end{equation}
In the $U$-gauge $\zeta$ describes instead the curvature fluctuations in the hypersurfaces 
where the total energy density is uniform. In the limit of large length-scales there seem to be 
no difference between $\zeta$ and ${\mathcal R}$. However, thanks to Eqs. (\ref{two22})--(\ref{two23}) the second-order equation obeyed by $\zeta$ is far more involved than Eqs. (\ref{one1}) and (\ref{one6}) even if the two equations coincide in the $k\tau\ll 1$ limit. The decoupled equation for $\zeta$  is formally non-local since it contains the inverse of the function $1 + k^2/[ 3({\mathcal H}^2 - {\mathcal H}^{\prime})]$. To lowest order in  $k \tau < 1$ we have that $f(k,\tau) \to 1$: in this limit $\zeta$ and ${\mathcal R}$ evolve at the same rate. In Fourier space the evolution of $\zeta_{\vec{k}}$ 
can then be written as:
\begin{eqnarray}
\zeta_{\vec{k}}^{\prime\prime} + {\mathcal H} [ 1 + f_{\vec{k}} + 3 c_{st}^2 (f_{\vec{k}} -1)] \zeta_{\vec{k}}^{\prime} 
+ k^2 \, c_{st}^2 \biggl[ 1 - \frac{1 + 3 c_{st}^2 }{3\, c_{st}^2 } f_{\vec{k}} \zeta_{\vec{k}} \biggr] \zeta_{\vec{k}} = {\mathcal S}_{\zeta}, \quad f_{\vec{k}}(\tau) = \frac{1}{1 + \frac{k^2}{ 3({\mathcal H}^2 - {\mathcal H}^{\prime})}},
\label{two24}
\end{eqnarray}
where ${\mathcal S}_{\zeta}$ is given by
\begin{equation}
{\mathcal S}_{\zeta} = \biggl(\Sigma_{\vec{k}} - \frac{{\mathcal H} \Pi_{t}}{\rho_{t} + p_{t}}\biggr)^{\prime} +
 {\mathcal H} [ 1 + f_{\vec{k}} + 3 c_{st}^2 (f_{\vec{k}} -1)] \biggl(\Sigma_{\vec{k}} - \frac{{\mathcal H} \Pi_{t}}{\rho_{t} + p_{t}}\biggr) - \frac{1}{{\mathcal H}} \nabla^2 \biggl( \Sigma_{{\mathcal R}} -  \frac{{\mathcal H} \Pi_{t}}{\rho_{t} + p_{t}}\biggr).
 \label{two25}
 \end{equation}
Given the expression of $f_{\vec{k}}(\tau)$, Eq. (\ref{two24}) is non-local. In the limit $k^2 \ll ({\mathcal H}^2 - {\mathcal H}^{\prime})$ Eqs. (\ref{two24}) and (\ref{two25}) are compatible with the evolution of ${\mathcal R}$ as implied by Eq. (\ref{two22}). Non-local terms can therefore be avoided with specific approximations which are are however 
unnecessary if the gauge-invariant evolution is studied and solved in terms of ${\mathcal R}$ and ${\mathcal R}^{\prime}$.
After having computed ${\mathcal R}$ and ${\mathcal R}^{\prime}$ the value of $\zeta$ can always be obtained 
from Eq. (\ref{two22}). We therefore conclude that if the effective anisotropic stresses are 
expressed in terms of  $\zeta_{\vec{k}}$ the evolution will necessarily involve non-local terms 
that are however absent in the approach suggested in this paper.

\renewcommand{\theequation}{4.\arabic{equation}}
\setcounter{equation}{0}
\section{Derivation in the uniform curvature gauge}
\label{sec4}
 Recalling Eqs. (\ref{onea10}), (\ref{onea11})--(\ref{onea12}) and (\ref{onea12bb}) the gauge-freedom can also be completely removed  in the coordinate system characterized by the following perturbed metric:
 \begin{equation}
\delta_{s}^{(1)} g_{00}(\vec{x},\tau) = 2 a^2(\tau)\, \phi(\vec{x},\tau), \qquad  \delta_{s}^{(1)} g_{0i}(\vec{x},\tau) = - a^2 V_{i}(\vec{x},\tau).
\label{three1}
\end{equation}
Even if we shall eventually set $V_{i} = \partial_{i} B$ it will be convenient, just for 
 the notational convenience, to write the general formulas in terms of $V_{i}$. In this section we shall therefore 
 repeat in the $U$-gauge all the steps leading to Eq. (\ref{two15}). The expression obtained in the $U$-gauge
 will still depend on ${\mathcal R}_{\vec{k}}$ and ${\mathcal R}_{\vec{k}}^{\prime}$ but it will be sharply 
 different from the expression of the $L$-gauge. This is what we meant in section\ref{sec1} when 
 introducing the concept of a gauge-invariant comparison of gauge-dependent results. 
 
\subsection{The effective anisotropic stress in terms of $\phi$ and $V_{i}$} 
In the $U$-gauge the effective anisotropic stress follows from Eq. (\ref{onea4})
with $X =U$:
\begin{equation}
\Pi_{i}^{\,\,(U) \,j} = \delta_{s}^{(2)} \, T_{i}^{\,\,(U)\, j} - \frac{1}{\ell_{P}^2} \delta_{s}^{(2)} {\mathcal G}_{i}^{\,\,(U)\, j}. 
\label{three2}
\end{equation}
Neglecting, as usual, the terms proportional to the trace we will have that 
\begin{eqnarray}
\delta_{s}^{(2)} {\mathcal G}_{i}^{\,\,(U)\, j} &=& - \frac{\phi}{a^2} \biggl(\partial_{i} \, V^{j\,\,\prime} + \partial^{j} V_{i}^{\prime} \biggr)
- \frac{\phi^{\prime}}{2 a^2} \biggl(\partial_{i} \, V^{j} + \partial^{j} V_{i} \biggr)
\nonumber\\
&-& 2 \frac{{\mathcal H} \phi}{a^2} \biggl(\partial_{i} \, V^{j} + \partial^{j} V_{i} \biggr) + \frac{1}{a^2} \partial_{i} \phi \partial^{j} \phi - \frac{2 {\mathcal H}}{a^2} \, V^{j} \, \partial_{i} \phi,
\label{three3}\\
\delta_{s}^{(2)} T_{i}^{\,\,(U)\, j} &=& - \frac{4 {\mathcal H}}{a^4 \, \ell_{P}^4\, (\rho_{t} + p_{t})} \partial_{i} \phi \biggl[ {\mathcal H} 
\partial^{j} \phi + \biggl({\mathcal H}^2 - {\mathcal H}^{\prime}\biggr) V^{j} \biggr].
\label{three4}
\end{eqnarray}
Inserting Eqs. (\ref{three3}) and (\ref{three4}) into Eq. (\ref{three2}) the explicit expression of the effective anisotropic stress is therefore given by:
\begin{eqnarray}
\Pi_{i}^{\,\,(U) \,j}(\vec{x},\tau) &=& \frac{1}{a^2 \ell_{P}^2} \biggl\{  \phi\biggl(\partial_{i} \, V^{j\,\,\prime} + \partial^{j} V_{i}^{\prime} \biggr)
+ \frac{\phi^{\prime}}{2} \biggl(\partial_{i} \, V^{j} + \partial^{j} V_{i} \biggr)
\nonumber\\
&+& 2 {\mathcal H} \phi  \biggl(\partial_{i} \, V^{j} + \partial^{j} V_{i} \biggr) - \partial_{i} \phi \partial^{j} \phi + 2 {\mathcal H} \, V^{j} \, \partial_{i} \phi,
\nonumber\\
 &-& \frac{4 {\mathcal H}}{a^2 \, \ell_{P}^2\, (\rho_{t} + p_{t})} \partial_{i} \phi \biggl[ {\mathcal H} 
\partial^{j} \phi + \biggl({\mathcal H}^2 - {\mathcal H}^{\prime}\biggr) V^{j} \biggr]\biggr\}.
\label{three5}
\end{eqnarray}
Recalling that the total spatial derivatives do not contribute once 
projected on the tensor polarizations (see Eq. (\ref{one5}) and discussion thereafter), Eq. (\ref{three5}) can be 
finally expressed as 
\begin{equation}
\Pi_{i}^{\,\,(U) \,j}(\vec{x},\tau)  = \frac{1}{a^2 \ell_{P}^2 } \biggl[ 2 \phi\, \partial_{i} \, \partial^{j} B^{\prime} + \phi^{\prime} 
\partial_{i} \partial^{j} B + 4 {\mathcal H}\, \phi\, \partial_{i} \, \partial^{j} B + \frac{3 {\mathcal H}^2 - {\mathcal H}^{\prime}}{{\mathcal H}^2 - {\mathcal H}^{\prime}} \,\phi \,\partial_{i} \partial^{j} \phi \biggr],
\label{three6}
\end{equation}
where we used that $V_{i} = \partial_{i} B$. In Fourier space the relation connecting $B_{\vec{k}}$, $\phi_{\vec{k}}$ and the 
scalar anisotropic stress is given by:
\begin{equation}
B_{\vec{k}}^{\prime} + 2 {\mathcal H} B_{\vec{k}} = - \phi_{\vec{k}} + \Delta_{\vec{k}}.
\label{three7}
\end{equation}
so that Eq. (\ref{three6}) becomes, in Fourier space,
\begin{eqnarray}
\Pi_{i}^{\,\,(U) \,j}(\vec{q},\tau) &=&  - \frac{1}{(2\pi)^{3/2}} \int e^{i \, \vec{q}\cdot\vec{x}} \, \Pi_{i}^{\,\,(U) \,j}(\vec{x},\tau)
\nonumber\\
&=&- \frac{1}{(2\pi)^{3/2}\, a^2\, \ell_{P}^2} \int\, d^{3} k\,\, k_{i}\, k^{j} \biggl[ \phi_{\vec{q} - \vec{k}} \Delta_{\vec{k}} 
+ \phi_{\vec{k}} \Delta_{\vec{q} - \vec{k}} + \biggl(\frac{{\mathcal H}^2 + {\mathcal H}^{\prime}}{{\mathcal H}^2 - {\mathcal H}^{\prime}}\biggr) \phi_{\vec{k}} \, \phi_{\vec{q} -\vec{k}}
\nonumber\\
&+& \frac{1}{2} \biggl( \phi_{\vec{q} - \vec{k}}^{\prime} \, B_{\vec{k}} + \phi_{\vec{k}}^{\prime} B_{\vec{q} - \vec{k}} \biggr) 
\biggl].
\label{three8}
\end{eqnarray}
Equation (\ref{three8}) is the $U$-gauge analog of Eq. (\ref{two13}) which is instead valid in the $L$-gauge.
It is however clear that Eqs. (\ref{three8}) and (\ref{two13}) are not comparable in any way since 
the pivotal variables of each gauge obey a different set of equations. Note, incidentally, 
that the variable $\phi$ appearing in Eq. (\ref{two13}) is defined in the $L$-gauge 
whereas Eq. (\ref{three8}) holds in the $U$-gauge where 
the variable $\phi$ evolves in a different way.

\subsection{The effective anisotropic stress in terms of ${\mathcal R}$ and ${\mathcal R}^{\prime}$}
As it happens in the $L$-gauge  the pivotal variables of the $U$-gauge are univocally related 
to  the curvature perturbations on comoving orthogonal hypersurfaces:
\begin{eqnarray}
\phi_{\vec{k}} &=& - \frac{{\mathcal H}^2 - {\mathcal H}^{\prime}}{{\mathcal H}^2} \, {\mathcal R}_{\vec{k}},
\label{three9}\\ 
B_{\vec{k}} &=& -  \frac{({\mathcal H}^2 - {\mathcal H}^{\prime})}{{\mathcal H}^2\, k^2\, c_{st}^2 } \, 
\biggl( {\mathcal R}_{\vec{k}}^{\prime} - \Sigma_{\vec{k}}\biggr).
\label{three10}
\end{eqnarray}
Equations (\ref{three9}) and (\ref{three10}) are the $U$-gauge analog of Eqs. (\ref{two2})--(\ref{two3}) and (\ref{two14}).  Inserting  Eqs. (\ref{three9}) and (\ref{three10}) 
into Eq. (\ref{three8}) the effective anisotropic stress becomes:
\begin{eqnarray}
\Pi_{i\,j}^{\,\,(U)}(\vec{q},\tau) &=& - \frac{({\mathcal H}^2 - {\mathcal H}^{\prime})^2}{(2 \pi)^{3/2}  \,a^2 \,\ell_{P}^2 \, {\mathcal H}^4}  \, \int d^{3} k  \, k_{i} \, k_{j} \biggl\{  \frac{{\mathcal H}^2 + {\mathcal H}^{\prime}}{{\mathcal H}^2 - {\mathcal H}^{\prime}} {\mathcal R}_{\vec{k}} {\mathcal R}_{\vec{q} - \vec{k}} 
\nonumber\\
&+& \frac{k^2 + |\vec{q} - \vec{k}|^2 }{2 \, c_{st}^2 \, k^2\, |\vec{q}- \vec{k}|^2 } {\mathcal R}_{\vec{k}}^{\prime} {\mathcal R}_{\vec{q} - \vec{k}}^{\prime} + \frac{3}{2} {\mathcal H} ( w - c_{st}^2) \biggl[ \frac{{\mathcal R}_{\vec{k}}^{\prime} {\mathcal R}_{\vec{q} - \vec{k}}}{c_{st}^2 \, k^2} + 
\frac{{\mathcal R}_{\vec{q} - \vec{k}}^{\prime} {\mathcal R}_{\vec{k}}}{c_{st}^2 \, |\vec{q} - \vec{k}|^2} \biggr] 
\nonumber\\
&-& \frac{\Sigma_{\vec{q} - \vec{k}} [ {\mathcal R}_{\vec{k}}^{\prime} + 3 {\mathcal H} ( w- c_{st}^2) {\mathcal R}_{\vec{k}}]}{2 c_{st}^2 \, k^2}  - \frac{\Sigma_{\vec{k}} [ {\mathcal R}_{\vec{q} - \vec{k}}^{\prime} + 3 {\mathcal H} ( w- c_{st}^2) {\mathcal R}_{\vec{q}- \vec{k}}]}{2\,c_{st}^2\,, |\vec{q} - \vec{k}|^2}
\nonumber\\
&-& \frac{{\mathcal H}^2}{ ({\mathcal H}^2 + {\mathcal H}^{\prime})} ( {\mathcal R}_{\vec{q} - \vec{k}} \Delta_{\vec{k}}
+ {\mathcal R}_{\vec{k}} \Delta_{\vec{q} - \vec{k}}) \biggl\}.
\label{three11}
\end{eqnarray}
Equation  (\ref{three11}) is the $U$-gauge analog of Eq. (\ref{two15}). Since ${\mathcal R}_{\vec{k}}$ and ${\mathcal R}_{\vec{k}}^{\prime}$ obey Eq. (\ref{one6}) the results of Eqs. (\ref{two15}) and Eq. (\ref{three11}) can be 
directly compared since they are expressed in terms of the same set of gauge-invariant variables. 
This comparison will be explicitly illustrated in the following section. 
It is finally rather easy to verify that Eq. (\ref{one6}) can be directly derived in the $U$-gauge. This 
step will be omitted here since it mirrors exactly the analysis of the $L$-gauge. The procedure is to derive both sides of Eq. (\ref{three10}) and to eliminate $B_{\vec{k}}^{\prime}$ by first using Eq. (\ref{three7}). This step will lead to a dependence on $\phi_{\vec{k}}$ and $B_{\vec{k}}$ that will be
eliminated thanks to Eqs. (\ref{three9}) and (\ref{three10});  Eq. (\ref{one6}) will then be recovered. This 
proves that, unlike the evolution equations of a each particular gauge, both Eqs. (\ref{one1}) and (\ref{one6}) 
are invariant under infinitesimal coordinate transformations of the type introduced in Eq. (\ref{onea10aa}).

\renewcommand{\theequation}{5.\arabic{equation}}
\setcounter{equation}{0}
\section{Gauge-invariant comparison of  gauge-dependent results}
\label{sec5}
Equations (\ref{two15}) and (\ref{three11})  have been derived in two different gauges but are 
expressed in terms of the {\em same} gauge-invariant variables  obeying Eq. (\ref{one6}).
This is the operational definition of the strategy introduced in section \ref{sec1} namely a gauge-invariant comparison 
of the gauge-dependent results. This apparent oxymoron emphasizes  that the results obtained in different 
coordinate systems  can be compared in a physically meaningful way 
only by expressing the gauge-dependent results in terms of the gravitating 
normal modes of the system. Since this logic has never been used before, we
intend to illustrate the power of our method by studying the limits of the effective anisotropic stresses 
when the typical wavelengths are either smaller or larger than the sound horizon $r_{s}(\tau)$.

\subsection{Wavelengths inside the sound horizon}
Let us first consider the simplest physical case where the non-adiabatic pressure fluctuations vanish, 
the scalar anisotropic stress is absent and the wavelengths of the scalar phonons are sufficiently small; in formulas
\begin{equation}
\Sigma_{\vec{k}} \to 0, \qquad \Delta_{\vec{k}} \to 0, \qquad k^2 c_{st}^2 \gg \biggl|\frac{z_{t}^{\prime\prime}}{z_{t}}\biggr|.
\label{five0a}
\end{equation}
In the situation of Eq. (\ref{five0a}) we have that Eq. (\ref{one6}) becomes
\begin{equation}
q_{\vec{k}}^{\prime\prime} + k^2 \, c_{st}^2 q_{\vec{k}} =0, \qquad q_{\vec{k}} = z_{t} \, {\mathcal R}_{\vec{k}}
\label{five0b}
\end{equation}
It is relevant to stress that, in this regime, the solution of Eqs. (\ref{one6}) and (\ref{five0b}) follows from Wentzel-Kramers-Brillouin (WKB) approximation without specifying the background evolution and it is given by:
\begin{equation}
{\mathcal R}_{\vec{k}}(\tau) = \frac{C_{\vec{k}}}{z_{t}\,\sqrt{2 \,k \, c_{st}}} \,\cos{[ k\, r_{s}(\tau)]} + \frac{D_{\vec{k}}}{z_{t}\,\sqrt{2 \,k \, c_{st}}}\,\sin{[ k\, r_{s}(\tau)]}.
\label{five1}
\end{equation}
Equation (\ref{five1}) is a WKB solution of Eq. (\ref{five0b}) provided $c_{st}^{\prime}/(2 c_{st}) < k \, c_{st}$;
in Eq. (\ref{five1}) $C_{\vec{k}}$ and $D_{\vec{k}}$ are two constants (possibly determined from the boundary conditions) and $r_{s}(\tau)$ defines the {\em sound horizon}:
\begin{equation}
r_{s}(\tau) = \int_{\tau_{i}}^{\tau} \, c_{st}(\tau) \, d\tau.
\label{five2}
\end{equation}
The wavelengths satisfying Eq. (\ref{five1}) will be said to be inside the sound horizon (i.e. $ k\,r_{s}(\tau) \gg 1$).

We are now going to consider separately the limits of Eqs. (\ref{two15}) and (\ref{three11}) {\em inside the sound horizon}.
For the sake of illustration we shall first consider the case of Eq. (\ref{five1}) and then 
comment on the main differences when $\Sigma_{\vec{k}} \neq 0$ and $\Delta_{\vec{k}} \neq 0$.
From Eq. (\ref{two15})  the expression of the effective anisotropic stress 
in the $L$-gauge becomes:
\begin{eqnarray}
\Pi_{i\, j}^{(L)}(\vec{q}, \tau) &=& - \frac{2 ({\mathcal H}^2 - {\mathcal H}^{\prime})}{(2\pi)^{3/2}\,\ell_{P}^2\, a^2(\tau)\,{\mathcal H}^2 } \int\, d^{3}k\, \, k_{i} \,\, k_{j} \biggl\{ {\mathcal R}_{\vec{k}} \, {\mathcal R}_{\vec{q} - \vec{k}} + \frac{{\mathcal H}^2 - {\mathcal H}^{\prime}}{{\mathcal H}} \biggl[ \frac{{\mathcal R}_{\vec{k}} \, {\mathcal R}_{\vec{q} - \vec{k}}^{\prime}}{c_{st}^2  \, |\vec{q} - \vec{k}|^2} 
\nonumber\\
&+& \frac{{\mathcal R}_{\vec{k}}^{\prime} \, {\mathcal R}_{\vec{q} - \vec{k}} }{c_{st}^2  \, k^2}\biggr] + 
\frac{( 2 {\mathcal H}^2 - {\mathcal H}^{\prime}) ({\mathcal H}^2 - {\mathcal H}^{\prime})}{{\mathcal H}^2 \, c_{st}^4 \, k^2 \, |\vec{q}- \vec{k}|^2 } \, {\mathcal R}_{\vec{k}}^{\prime}\,  {\mathcal R}_{\vec{q} - \vec{k}}^{\prime}\biggr\}.
\label{five3}
\end{eqnarray}
From  Eq. (\ref{five1}) inside the sound horizon,  the curvature perturbations and their derivatives are approximately related as 
\begin{equation}
{\mathcal R}_{\vec{k}}^{\prime} \simeq k \,c_{st}\, {\mathcal R}_{\vec{k}}\biggl[ 1 + {\mathcal O}\biggl(\frac{a\, H}{k\, c_{st}}\biggr) + \, .\, .\, .\biggr],
\label{five3a}
\end{equation}
and this relation holds in spite of the details of the underlying background geometry;  the ellipses in Eq. (\ref{five3a}) denote the higher-order corrections that are always negligible for $k \, c_{s t} \gg H \, a$; an analog expansion holds in the case ${\mathcal R}_{\vec{q} - \vec{k}}^{\prime}$ when 
$|\vec{q} - \vec{k}| \, c_{st} \gg a\, H$. Inserting Eq. (\ref{five3a}) into Eq. (\ref{five3}) we get the following 
result:
\begin{eqnarray}
\Pi_{ij}^{\,(L)}(\vec{q}, \tau)  &=& - \frac{2 ({\mathcal H}^2 - {\mathcal H}^{\prime})}{(2\pi)^{3/2}\,\ell_{P}^2\, a^2(\tau)\,{\mathcal H}^2 } \int\, d^{3}k\, \, k_{i} \,\, k_{j} {\mathcal R}_{\vec{k}} \, {\mathcal R}_{\vec{q} - \vec{k}}\, \biggl\{ 1+ ({\mathcal H}^2 - {\mathcal H}^{\prime})  \frac{( k+ |\vec{q} - \vec{k}|)}{c_{st}  \, {\mathcal H}\, k\, |\vec{q} - \vec{k}|}\nonumber\\
&+& \frac{( 2 {\mathcal H}^2 - {\mathcal H}^{\prime})}{{\mathcal H}^2 \, c_{st}^2 \, k \, |\vec{q}- \vec{k}| } + \, .\, .\,. \biggr\},
\label{five4}
\end{eqnarray}
where, as in Eq. (\ref{five3a}),  the ellipses stand for the higher-order contributions.  The first term at the right hand side of Eq. (\ref{five4}) dominates in the limit $k\, c_{st} \gg H a$  while the two remaining contributions are 
of higher order. For short the range of validity of Eq. (\ref{five4}) can be dubbed as $k \, c_{st} \, \tau \gg 1$, $|\vec{q} - \vec{k}| \, c_{st}\, \tau \gg 1$ with $(k \, c_{st} \, \tau)/(|\vec{q} - \vec{k}| \, c_{st}\, \tau ) \to 1$ since ${\mathcal H} = a \, H \sim 1/\tau$. Note, however, that this is just some kind of shorthand notation that does not imply the choice of a specific background as it happens in gauge-dependent and background-dependent studies. 

The same analysis leading to Eq. (\ref{five4}) can be repeated in the $U$- gauge. More 
specifically we have that for $\Sigma_{\vec{k}} \to 0$ and $\Delta_{\vec{k}} \to 0$  Eq. (\ref{three11}) becomes:
\begin{eqnarray}
\Pi_{ij}^{(U)}(\vec{q}, \tau) &=& - \frac{({\mathcal H}^2 - {\mathcal H}^{\prime})^2}{(2\pi)^{3/2} \,\ell_{P}^2\, a^2(\tau)\, {\mathcal H}^4} \int d^{3} k\,\, k_{i}\,k_{j} \,\biggl\{  \biggl( \frac{{\mathcal H}^2 + {\mathcal H}^{\prime}}{{\mathcal H}^2 - {\mathcal H}^{\prime}}\biggr) {\mathcal R}_{\vec{k}} \,{\mathcal R}_{\vec{q} - \vec{k}} 
\nonumber\\
&+& \frac{3}{2} {\mathcal H} (w - c_{st}^2) \biggl[ \frac{{\mathcal R}_{\vec{q} - \vec{k}}^{\prime} \, {\mathcal R}_{\vec{k}}}{|\vec{q} - \vec{k}|^2 \, c_{st}^2} + \frac{{\mathcal R}_{\vec{q} - \vec{k}}\, {\mathcal R}_{\vec{k}}^{\prime}}{k^2 \, c_{st}^2} \biggr] +  \frac{k^2 + |\vec{q} - \vec{k}|^2}{2\,c_{st}^2 \, |\vec{q} - \vec{k}|^2\,k^2} {\mathcal R}_{\vec{k}}^{\prime}  
{\mathcal R}_{\vec{q} -\vec{k}}^{\prime} \biggr\}.
\label{five5}
\end{eqnarray}
At this point it is important to recall that Eqs. (\ref{one6}) and (\ref{five0b}) are both gauge-invariant: they are therefore the same in any coordinate system. Equation (\ref{five3a}) can then be inserted into Eq. (\ref{five5}) 
so that, in the limits $k \, c_{st} \, \tau \gg 1$, $|\vec{q} - \vec{k}| \, c_{st}\, \tau \gg 1$ with $(k \, c_{st} \, \tau)/(|\vec{q} - \vec{k}| 
\, c_{st}\, \tau ) \to 1$ the same steps leading to Eq. (\ref{five4}) leads, 
in the case of Eq. (\ref{five5}), to the following result:
\begin{eqnarray}
\Pi_{i j}^{(U)}(\vec{q},\tau) &=& - \frac{({\mathcal H}^2 - {\mathcal H}^{\prime})^2}{(2\pi)^{3/2} \, \ell_{P}^2\, {\mathcal H}^4 a^2}
\,\, \int d^{3} k \,\, k_{i} \, k_{j} {\mathcal R}_{\vec{k}} \, {\mathcal R}_{\vec{q} - \vec{k}} \biggl\{ 1 + \frac{{\mathcal H}^2 
+ {\mathcal H}^{\prime}}{{\mathcal H}^2 - {\mathcal H}^{\prime}} 
\nonumber\\
&+& \frac{3}{2} {\mathcal H} \frac{( w - c_{st}^2)\,(k + |\vec{q} - \vec{k}|)}{c_{st}\,|\vec{q} - \vec{k}| \, k}  + \,.\,.\,.\biggr\}.
\label{five6}
\end{eqnarray}

The direct comparison of Eqs. (\ref{five4}) and (\ref{five6})  demonstrates
that the leading terms of both expansions are the same. Therefore, as long as 
the  wavelengths of the gravitating normal modes are shorter than the sound horizon the
anisotropic stresses will coincide up to subleading corrections:
\begin{equation}
\Pi_{ij}^{(L)}(\vec{q}, \tau) = \Pi_{ij}^{(U)}(\vec{q},\tau) + {\mathcal O}\biggl(\frac{a\, H}{k \, c_{st}}\biggr) + {\mathcal O}\biggl(\frac{a\, H}{|\vec{q} - \vec{k}| \, c_{st}}\biggr)  + {\mathcal O}\biggl(\frac{a^2\, H^2}{k \, |\vec{q} - \vec{k}| \, c_{st}^2}\biggr) +\,.\,.\,.
\label{five7}
\end{equation}
So far we considered the case $\Sigma_{\vec{k}} \to 0$ and $\Delta_{\vec{k}} \to 0$. To avoid a repetitive discussion we shall only mention the main differences arising in the case $\Sigma_{\vec{k}} \neq 0$ and $\Delta_{\vec{k}} \neq 0$. Equation (\ref{five1}) must be replaced by the following  equation
\begin{equation}
{\mathcal R}_{\vec{k}}(\tau) = {\mathcal R}^{(1)}_{\vec{k}}(\tau) + \int_{\tau_{i}}^{\tau} \, d\xi \, 
G_{{\mathcal R}}[ q( \xi - \tau)] \, {\mathcal S}_{\mathcal R}(\xi),
\label{five8}
\end{equation}
where ${\mathcal R}^{(1)}_{\vec{k}}(\tau)$ is the solution of the homogeneous equation 
given in (\ref{five1}) while 
 $G_{{\mathcal R}}[ q( \xi - \tau)]$ and ${\mathcal S}_{\mathcal R}(\xi)$ are defined as
 \begin{equation}
 G_{{\mathcal R}}[ q( \xi - \tau)] = - \frac{z_{t}(\xi)}{q \, c_{st}\, z_{t}(\tau)} \sin{[q(\xi - \tau)]},\, 
 \qquad S_{{\mathcal R}}(\xi) = \Sigma_{\vec{k}}^{\prime} + 2 \frac{z_{t}^{\prime}}{z_{t}} \Sigma_{\vec{k}} + 
 \frac{3 a^4}{z_{t}^2} \Pi_{\vec{k}}.
 \label{five9}
 \end{equation}
 Equation (\ref{five8}) must then be inserted into Eqs. (\ref{two15}) and (\ref{three1}). Using the 
 analog of Eq. (\ref{five3a}) and neglecting all the terms that are subheading inside the sound horizon 
 the result of Eq. (\ref{five7}) can be recovered.
 
To reach the previous conclusion it is relevant to appreciate that inside the sound 
 horizon the scalar anisotropic stress is more suppressed than the curvature inhomogeneities. 
 Since this is a relevant aspect it seems appropriate to justify it in more detail in the 
 simplest situation where the scalar anisotropic stress comes exclusively 
 from the neutrino sector. Let us therefore write the coupled evolution of the 
 curvature inhomogeneities and of the scalar anisotropic stress in the case of a radiation-dominated 
background. Equations (\ref{one6}) and  (\ref{one19}) read
\begin{eqnarray}
&& \frac{d^2 {\mathcal R}_{\vec{k}}}{d y^2} + \frac{2}{y} \frac{d {\mathcal R}_{\vec{k}}}{d y} + c_{st}^2\, {\mathcal R}_{\vec{k}} = \frac{R_{\nu}}{y} \biggl( \frac{d \sigma_{\nu}}{d y} + \frac{2}{y} \sigma_{\nu} \biggr),
 \label{five10}\\
 && \frac{d^3 \sigma_{\nu}}{d y^3} + \biggl(\frac{6}{7} + \frac{8 R_{\nu}}{5 y^2} \biggr)\frac{d \sigma_{\nu}}{d y} - 16 \frac{R_{\nu}}{y^3}\sigma_{\nu}
 + \frac{16 }{5 y^2 } \biggl( 3 \frac{d {\mathcal R}}{d y} +  \frac{y {\mathcal R}}{3}  \biggr) =0,
\label{five11}
\end{eqnarray}
where  $c_{st} = 1/\sqrt{3}$ and $ y = k\tau$; to derive Eqs. (\ref{five10}) and (\ref{five11}) we assumed 
$\Pi_{\vec{k}} = (\rho_{\nu} + p_{\nu}) \sigma_{\nu}$. Inside the sound horizon the dominant solution
of Eqs. (\ref{five10}) and (\ref{five11}) reads:
\begin{equation}
{\mathcal R}_{\vec{k}}(\tau) = \overline{{\mathcal R}}(\vec{k}) \frac{\sin{c_{st} \, y}}{ c_{st} \, y}, \qquad 
\sigma_{\nu}(\vec{k}, \tau) = \overline{\sigma}(\vec{k})\frac{\sin{c_{st} \, y}}{ |c_{st} \, y|^2},
\label{five12}
\end{equation}
where $\overline{\sigma}(\vec{k}) = (112/55) \overline{{\mathcal R}}(\vec{k})$. Therefore, as anticipated, 
inside the sound horizon (i.e. for $c_{st} y \gg 1$) the scalar anisotropic stress is always more suppressed 
in comparison with the curvature inhomogeneities. 

\subsection{Wavelengths outside the sound horizon}
So far we investigated the effective anisotropic stresses when the corresponding 
wavelengths are shorter than the sound horizon. We shall now consider the opposite limit 
where the  wavelengths are larger than the sound horizon i.e.
\begin{equation}
k\,r_{s}(\tau) \ll 1, \qquad k^2 \ll \biggl| \frac{z_{t}^{\prime\prime}}{z_{t}}\biggr|.
\label{five13}
\end{equation} 
To investigate the limit of Eq. (\ref{five13}) we first rewrite Eq. (\ref{one6}) in the following form:
  \begin{equation}
 \partial_{\tau} \biggl[ z_{t}^2 \biggl( {\mathcal R}_{\vec{k}}^{\prime} - \Sigma_{\vec{k}} \biggr) \biggr] = - k^2 c_{st}^2 \, z_{t}^2 \, {\mathcal R}_{\vec{k}} + 3 a^4 \, \Pi_{\vec{k}}.
 \label{five15}
 \end{equation}
Equation (\ref{five15}) has the same content of Eq. (\ref{one6}) but it can be easily transformed into an integral equation which 
will be easier to handle in this situation:
\begin{equation}
\biggl({\mathcal R}_{\vec{k}}^{\prime} - \Sigma_{\vec{k}} \biggr) = \biggl(\frac{z_{ex}}{z_{t}}\biggr)^2 \,\biggl({\mathcal R}_{\vec{k}}^{\prime} - \Sigma_{\vec{k}} \biggr)_{ex} - k^2 c_{st}^2 \int_{\tau_{ex}}^{\tau} z_{t}^2(\tau_{1}) {\mathcal R}_{\vec{k}}(\tau_{1}) 
+ 3 \int_{\tau_{ex}}^{\tau} a^4(\tau_{1}) \, \Pi_{\vec{k}}(\tau_{1}) d \tau_{1}.
\label{five16}
\end{equation}
In Eq. (\ref{five16}) $\tau_{ex}$ denotes the time at which the given scale exits the sound horizon  (i.e. $k c_{st} \tau_{ex} \simeq 1$); during inflation $z_{t} \to z_{\varphi} = a\, \varphi^{\prime}/{\mathcal H}$ (where $\varphi$ is the inflaton) so that $k \tau_{ex} \simeq 1$ and the sound horizon coincides, in practice, with the Hubble radius. 
For wavelengths larger than the sound horizon the scalar anisotropic stress is negligible 
with respect to ${\mathcal R}_{\vec{k}}$; this is what happens in the case of the concordance 
paradigm both in the case of the standard adiabatic mode and in the case of the other entropic modes \cite{TWENTYONEc,TWENTYTWOh,TWENTYTWOi}.

When the typical wavelengths are larger than the sound horizon the evolution 
of the curvature perturbations follows from Eq. (\ref{five16}). Now the idea will be to insert Eq. (\ref{five16}) both into Eq. (\ref{two15}) and into Eq. (\ref{three11}); at the very end the two expressions shall be compared. In general 
terms the effective anisotropic stresses in the $L$-gauge and in the $U$-gauge are expressible as:
\begin{equation}
\Pi_{ij}^{(X)}(\vec{q},\tau) = - \frac{1}{(2 \pi)^{3/2} \,a^2(\tau)\, \ell_{P}^2} \int d^{3} k \, k_{i} \, k_{j} \, A^{(X)}(\vec{k}, \vec{q}, \tau) \biggl[ 1  +   {\mathcal O}\biggl(\frac{k \, c_{st}}{a H}\biggr) + 
{\mathcal O}\biggl(\frac{|\vec{q} - \vec{k}| \, c_{st}}{a H}\biggr) + .\,.\,.\biggr],
\label{five17}
\end{equation}
where $X= L,\, U$; the expansion of Eq. (\ref{five17}) holds when the corresponding 
wavelengths are larger than the sound horizon (i.e. $k \, c_{st} < a\, H$ and $|\vec{q} - \vec{k}| \, c_{st} < a\, H$) 
and $A^{(X)}(\vec{k}, \vec{q}, \tau)$ is the leading term in the expansion obtained after the insertion 
of Eq. (\ref{five16}) into  Eqs. (\ref{two15}) and (\ref{three11}).
The explicit form of the leading contribution in the $L$-gauge reads:
\begin{equation}
A^{(L)}(\vec{k}, \vec{q}, \tau) = 2 \frac{(2 {\mathcal H}^2 - {\mathcal H}^{\prime})({\mathcal H}^2 - {\mathcal H}^{\prime})^2}{{\mathcal H}^4\, \, k^2 \, |\vec{q} -\vec{k}|^2 \, c_{st}^4} \biggl(\frac{z_{ex}}{z_{t}}\biggr)^{4} \, \biggl({\mathcal R}_{\vec{k}}^{\,\,\prime} - \Sigma_{\vec{k}}\biggr)_{ex} \,\,  \biggl({\mathcal R}_{\vec{q} - \vec{k}}^{\,\,\prime}- \Sigma_{\vec{q} - \vec{k}}\biggr)_{ex},
\label{five18}
\end{equation}
In the $U$-gauge the leading contribution implies instead:
\begin{eqnarray}
A^{(U)}(\vec{k}, \vec{q}, \tau) &=& \frac{({\mathcal H}^2 - {\mathcal H}^{\prime})^2}{2 {\mathcal H}^4}
\biggl(\frac{k^2 + |\vec{q} - \vec{k}|^2|}{k^2 \, |\vec{q} - \vec{k}|^2 \, c_{st}^2} \biggr) 
\nonumber\\
&\times& \biggl[ \Sigma_{\vec{k}} + \biggl(\frac{z_{ex}}{z_{t}}\biggr)^2 \biggl({\mathcal R}_{\vec{k}}^{\prime} - 
\Sigma_{\vec{k}} \biggr)_{ex} \biggr] \biggl[ \Sigma_{\vec{q}- \vec{k}} + \biggl(\frac{z_{ex}}{z_{t}}\biggr)^2 \biggl({\mathcal R}_{\vec{q}- \vec{k}}^{\prime} - 
\Sigma_{\vec{q} - \vec{k}} \biggr)_{ex} \biggr].
\label{five19}
\end{eqnarray}
The ratio between Eq. (\ref{five19}) and Eq. (\ref{five18}) is therefore the following:
 \begin{eqnarray}
 \frac{A^{(U)}(\vec{k}, \vec{q}, \tau)}{A^{(L)}(\vec{k}, \vec{q}, \tau)} &=& \frac{ ( k^2 + |\vec{q} - \vec{k}|^2) \, c_{st}^2}{4 ( 2 {\mathcal H}^2 - {\mathcal H}^{\prime})} 
\nonumber\\
&\times& \biggl[ 1 + \frac{\Sigma_{\vec{k}}}{({\mathcal R}_{\vec{k}}^{\prime} - \Sigma_{\vec{k}})_{ex}} \biggl(\frac{z_{t}}{z_{ex}}\biggr)^2 \biggr] \biggl[ 1 + \frac{\Sigma_{\vec{q} -\vec{k}}}{({\mathcal R}_{\vec{q}-\vec{k}}^{\prime} - \Sigma_{\vec{q} - \vec{k}})_{ex}} \biggl(\frac{z_{t}}{z_{ex}}\biggr)^2 \biggr].
 \label{five20}
\end{eqnarray}
From  Eq. (\ref{five20}) we see that 
 $A^{(U)}(\vec{k}, \vec{q}, \tau)/A^{(L)}(\vec{k}, \vec{q}, \tau)= {\mathcal O}(k^2 c_{st}^2 \tau^2)$ which is always smaller than $1$ when the corresponding wavelengths are larger than the sound horizon at the corresponding epoch. 

Let us therefore summarize the main conclusions reached so far. We started by suggesting in section \ref{sec1} a 
gauge-invariant comparison of gauge-dependent results. The novel idea of this comparison has been to 
express the effective anisotropic stresses directly in terms of the gravitating normal modes of the plasma
which obey the same evolution equation in any coordinate system. Inside the sound horizon 
the effective anisotropic stresses computed in the $L$-gauge and in the $U$-gauge 
coincide to leading order and this conclusion is summarized by Eqs. (\ref{five4}), (\ref{five6}) and (\ref{five7}). 
For typical wavelengths larger than the sound horizon Eqs. (\ref{five18}), (\ref{five19}) and (\ref{five20}) 
imply instead that the anisotropic stresses are sharply different and that, in particular, the result 
in the $U$-gauge is much smaller than the one in the $L$-gauge. The obtained 
result suggest therefore the important conclusion that the effective anisotropic stresses 
are approximately gauge-invariant inside the sound horizon but sharply different outside of it.

\subsection{Extensions to more general  situations}

The conclusion reached so far holds in a rather general situation and, in particular, 
when the evolution of the curvature inhomogeneities obeys Eq. (\ref{one6}).
Even more general situations are described by a similar equation where 
however the source terms have a different expression. In particular 
further sources of anisotropic stress (besides the fluid component) 
and further sources of entropy perturbations can be always rephrased in a form similar to the one 
of Eqs. (\ref{one7}) and (\ref{one8}). To substantiate this statement we can consider
the effect of the electric and magnetic fields on the scalar modes
The fluctuations of the energy density and the anisotropic stresses are both quadratic in the electric and magnetic fields and are defined as 
\begin{eqnarray}
&& \delta_{s} \rho_{B}(\vec{x},\tau) = \frac{B^2(\vec{x},\tau)}{4 \pi a^4}, \qquad \delta_{s} \rho_{E}(\vec{x},\tau) = \frac{E^2(\vec{x},\tau)}{4 \pi a^4},
\label{one11}\\
&& \Pi_{i}^{(B)\,\,j}= \frac{1}{4 \pi a^a} \biggl[B_{i}\,B^{\,\,j}- \frac{B^2}{3} \delta_{i}^{\,\,j} \biggr], \qquad 
\Pi_{i}^{(E)\,\,j}= \frac{1}{4 \pi a^a} \biggl[E_{i}\,E^{\,\,j}- \frac{BE^2}{3} \delta_{i}^{\,\,j} \biggr],
\label{one12}
\end{eqnarray}
where $\vec{E}$ and $\vec{B}$ denote the comoving electric and magnetic fields (see Ref. \cite{TWENTYTHREE} and discussion 
therein). 
Using the standard notations for the scalar components of the magnetic and electric anisotropic stresses 
\begin{equation}
\nabla^2 \Pi_{B}(\vec{x},\tau) = \partial_{i} \partial_{j} \Pi^{ij}_{(B)}(\vec{x},\tau), \qquad 
\nabla^2 \Pi_{E}(\vec{x},\tau) = \partial_{i} \partial_{j} \Pi^{ij}_{(E)}(\vec{x},\tau),
\label{one13}
\end{equation}
the generalized expression for $\overline{\Sigma}_{{\mathcal R}}(\vec{x},\tau)$ now becomes\footnote{Note that 
we used $\overline{\Sigma}_{{\mathcal R}}$ to distinguish it from $\Sigma_{{\mathcal R}}$ where the electromagnetic contribution 
is basent. }:
\begin{equation}
\overline{\Sigma}_{{\mathcal R}}(\vec{x},\tau) = - \frac{{\mathcal H}}{p_{t} + \rho_{t}} \delta p_{nad} + \frac{{\mathcal H}}{p_{t} + \rho_{t}} \biggl[ \biggl( c_{\mathrm{st}}^2 - \frac{1}{3}\biggr)(\delta_{s}\rho_{E} + \delta_{s}\rho_{B})  + \Pi_{t} + \Pi_{E} + \Pi_{B} \biggr].
\label{one14}
\end{equation}
Equation (\ref{one14}) generalizes the results of Eqs. (\ref{one6}) and (\ref{one7}) and can be used to compute 
the evolution of the curvature inhomogeneities in the presence of electromagnetic disturbances. The main 
observation we ought to make is that the form of Eq. (\ref{one6}) is exactly the same but, this time ${\mathcal S}_{{\mathcal R}}$
is replaced by $\overline{{\mathcal S}}_{{\mathcal R}}$:
\begin{equation}
\overline{{\mathcal S}}_{{\mathcal R}}(\vec{x},\tau) = \overline{\Sigma}_{\mathcal R}' + 2 \frac{z_{t}'}{z_{t}} \overline{\Sigma}_{\mathcal R}+ \frac{ 3 a^{4}}{z_{t}^2 } \biggl(\Pi_{t} + \Pi_{E} + \Pi_{B}\biggr).
\label{one14a}
\end{equation}
This also means, for instance, that the results of Eqs. (\ref{five18}), (\ref{five19}) and (\ref{five20}) can be 
easily deduced also in the presence of electromagnetic components by simply replacing $\Sigma_{\vec{k}}$ 
with $\overline{\Sigma}_{\vec{k}}$ and by redefining $\Pi_{t}$ as $\overline{\Pi}_{t} = \Pi_{t} + \Pi_{E} + \Pi_{B}$.

\renewcommand{\theequation}{6.\arabic{equation}}
\setcounter{equation}{0}
\section{The example of the concordance paradigm}
\label{sec6}
\subsection{Basic considerations}
The conclusions reached so far do not assume any specific background evolution.  It is however useful to corroborate 
the results obtained so far with the illustrative example of a radiation-dominated plasma. Most of the discussion 
could be conducted in terms of a generic sound speed but for the sake of concreteness 
we shall consider the situation 
\begin{equation}
c_{st}^2 = w = \frac{1}{2}, \qquad {\mathcal H} \, a = {\mathcal H}_{1} \,a_{1},\qquad  H \, a^2 =  H_{1} \,a^2_{1},
\label{six1}
\end{equation}
where ${\mathcal H} a$ as well as  $H \, a^2$ are constants throughout all the stages of the evolution. 
Furthermore we shall neglect both the non-adiabatic pressure fluctuations and the sources of scalar 
anisotropic stress (e.g. neutrinos). In the case (\ref{six1}) 
the scalar mode functions  can be computed in a closed form:
\begin{equation}
{\mathcal R}_{\vec{q}}(\tau) = \overline{{\mathcal R}}(\vec{q}\,) \, j_{0}(q \, c_{st}\, \tau), \qquad {\mathcal R}^{\prime}_{\vec{q}}(\tau) = - q\, c_{st} \,\overline{{\mathcal R}}(\vec{q}\,) \,j_{1}(q \, c_{st} \,\tau),
\label{six2}
\end{equation}
where $j_{0}(q \, c_{st} \,\tau)$ and $j_{1}(q \, c_{st} \,\tau)$ are spherical Bessel functions of 
zeroth- and first-order \cite{EX1,EX2}. To identify more easily the various different contributions in the effective anisotropic stress the sound speed  has been kept constant but generic in Eq. (\ref{six2}) (we shall eventually set $c_{st} \to 1/\sqrt{3}$ only at the very end). In Eq. (\ref{six2}) ${\mathcal R}(\vec{q}\,)$ represents a scalar random field, 
whose correlation function and the associated power spectrum are:
\begin{equation}
\langle \overline{{\mathcal R}}(\vec{q}\,) \, \overline{{\mathcal R}}(\vec{q}^{\,\,\prime}) \rangle = \frac{2 \pi^2}{q^3} \overline{P}_{{\mathcal R}}(q) \, \delta^{(3)}(\vec{q} + 
\vec{q}^{\,\,\prime}), \qquad \overline{P}_{{\mathcal R}}(q) = {\mathcal A}_{{\mathcal R}} \biggl(\frac{q}{q_{p}}\biggr)^{n_{s} -1}.
\label{six3}
\end{equation}
In Eq. (\ref{six3}) we used the standard normalizations 
where ${\mathcal A}_{{\mathcal R}}$ is the amplitude of the power spectrum at the pivot scale $q_{p} = 0.002\,\mathrm{Mpc}^{-1}$ corresponding to a frequency $\nu_{p} = 2 \pi q_{p} = 3\times10^{-18}$ Hz; in Eq. (\ref{six3}) $0.9 < n_{s} < 1$ denotes the scalar spectral index (see e.g. \cite{TWOa} and discussions therein). 
With the same notation employed in Eq. (\ref{six3}) the two-point function of a (solenoidal and traceless) tensor random field 
will be written as:
\begin{equation}
\langle \overline{h}_{i\,j}(\vec{q}\,) \,\overline{h}_{m\, n} (\vec{q}^{\,\,\prime}) \rangle = \frac{2 \pi^2}{q^3} \, {\mathcal S}_{i\,j\,m\,n}(\hat{q}) \,\overline{P}_{T}(q) \, \delta^{(3)}(\vec{q} + \vec{q}^{\,\,\prime}), \qquad \overline{P}_{T}(q) = {\mathcal A}_{T}  \biggl(\frac{q}{q_{p}}\biggr)^{n_{T}},
\label{six4}
\end{equation}
where ${\mathcal S}_{i\, j\, m\, n}(\hat{q})$ is related to the sum over the two tensor polarizations defined in Eq. (\ref{onea9}) 
and it is defined as:
\begin{equation}
{\mathcal S}_{i\, j\, m\, n}(\hat{q}) = \frac{1}{4} \biggl[p_{im}(\hat{q}) p_{jn}(\hat{q}) + p_{in}(\hat{q}) p_{jm}(\hat{q}) - p_{ij}(\hat{q}) p_{mn}(\hat{q})\biggr],\qquad p_{ij}(\hat{q}) = \delta_{ij} - \hat{q}_{i} \hat{q}_{j}. 
\label{six5}
\end{equation}
According to
the standard notations, ${\mathcal A}_{T} = r_{T} \, {\mathcal A}_{{\mathcal R}}$ 
is the amplitude of the tensor power spectrum at the same pivot scale used for the scalars.
The tensor to scalar ratio $r_{T}$ and the spectral index $n_{T}$ may be related by the so-called consistency relations (i.e. $n_{T} \simeq r_{T}/8$) but this point is not central for the present discussion.
In terms of the tensor random fields entering Eq. (\ref{six4}) the homogeneous solution of the equation of the tensor modes is
\begin{equation}
\overline{h}_{i\,j}(\vec{q},\tau) = \overline{h}_{i\,j}(\vec{q}\,) \, j_{0}(q\,\tau), \qquad H_{ij}(\vec{q},\tau) = \partial_{\tau} \overline{h}_{i\,j}(\vec{q},\tau) = - q \,\, \overline{h}_{i\,j}(\vec{q}\,) \, j_{1}(q\,\tau).
\label{six6}
\end{equation}

\subsection{Explicit evaluation of the effective anisotropic stresses in a radiation plasma}
Now the idea is, in short, the following:
\begin{itemize}
\item{} we are first going to insert Eqs. (\ref{six2}) and (\ref{six3}) into the exact expressions of the effective 
anisotropic stresses in the $L$-gauge and in the $U$-gauge obtained in 
 Eqs. (\ref{two15}) and (\ref{three11}) respectively;
\item{} then we shall compare the two exact expressions in the two physical limits 
when the wavelengths of the normal modes are either larger or smaller then the sound horizon;
\item{} finally we will compute the spectral energy density 
of the relic gravitons and explicitly evaluate the correction induced by the 
effective anisotropic stress.
\end{itemize}
When Eqs. (\ref{six2}) and (\ref{six3}) are inserted into Eqs. (\ref{two15}) and (\ref{three11})  the resulting 
expression of the effective anisotropic stress becomes:
 \begin{equation}
\Pi_{ij}^{(X)}(q,\tau) = - \frac{1}{(2\pi)^{3/2} \, \ell_{P}^2 \, a^2} \int d^{3} k\,\, k_{i} \, k_{j} \,\, \overline{{\mathcal R}}(\vec{k}) \, 
\overline{{\mathcal R}}(\vec{q} -\vec{k}) \, M^{(X)}(k \, c_{st}\,\tau, \, |\vec{q} - \vec{k}| \, c_{st} \, \tau).
\label{six7}
\end{equation}
The general expression of Eq. (\ref{six7}) is actually more general than the examples we are now describing; note, in particular, that $M^{(X)}(z, w)$ is  symmetric for $w\to z$ and $z\to w$. 
In the particular case of the radiation-dominated plasma discussed in Eq. (\ref{six1}) the exact expressions 
of $M^{(X)}(z,\,w)$ (for $X = L,\, U$) are:
\begin{eqnarray}
&& M^{(L)}(z,w) = 4 \biggl[ j_{0}(z) j_{0}(w) - 6 c_{st}\biggl(  \frac{j_{0}(z)j_{1}(w)}{w}  +\frac{  j_{1}(z) j_{0}(w)}{z} \biggr)
- \frac{54 c_{st}^2}{w \, z} j_{1}(w) j_{1}(z)\biggr],
\label{six8}\\
&& M^{(U)}(z,w) = 6 c_{st}^2 \, \biggl(\frac{z}{w} + \frac{w}{z} \biggr) j_{1}(w)  j_{1}(z).
\label{six9}
\end{eqnarray}
The variable $z$ appearing in Eqs. (\ref{six8}) and (\ref{six9}) 
has nothing to do with the variable $z_{t}$ appearing, for instance, in Eqs. 
(\ref{one1}) and (\ref{one6}). In the limits $z= k c_{st} \xi \gg 1$, $w= |\vec{q} - \vec{k}| c_{st} \xi \gg 1$ and $q\, c_{st} \xi \gg 1$, Eqs. (\ref{six8}) and (\ref{six9}) become:
\begin{equation}
M^{(L)}(z,\, w) \to \frac{ 4 \sin{z} \, \sin{w}}{w \, z} + .\,.\,.\,,\qquad M^{(U)}(z,\, w) \to \frac{ 12 c_{st}^2 \cos{z} \, \cos{w}}{w \, z}+ .\,.\,.\,.
\label{six10}
\end{equation}
Equations (\ref{six10}) apply when the wavelengths are all inside sound horizon (i.e. $k\, c_{st}/(a H) > 1$); however since $c_{st} \leq 1$ (and $k/(a\, H)> c_{st}^{-1}$) the wavelenghts are also inside the Hubble radius (i.e. $k/(a\, H)> 1$).  
It is important to appreciate that the results of Eq. (\ref{six10}) coincide exactly up o a phase and this because  $c_{st}= 1/\sqrt{3}$.
As we shall see later this phase will be immaterial for the final expression of the spectral energy density.
When the corresponding wavelengths are outside the sound horizon the asymptotic forms of Eqs. (\ref{six8}) and (\ref{six9}) are 
\begin{equation}
M^{(L)}(z, w) \to 4(6 c_{st}^2 - 4 c_{st} +1)+ .\,.\,.\,,,\qquad M^{(U)}(z, w) \to \frac{2}{3} c_{st}^2 [ z^2 + w^2]+ .\,.\,.\,,
\label{six11}
\end{equation}
respectively.

\subsection{The explicit expressions of the spectral energy density}
The solution of Eq. (\ref{onea9}) for $h_{\lambda}$ and $\partial_{\tau} h_{\lambda}$ is formally expressed in terms 
of the corresponding Green's functions $G[ q ( \xi- \tau)]$ and $\widetilde{\,G\,}[q ( \xi -\tau)]$:
\begin{eqnarray}
h_{\lambda}^{(X)}(\vec{q},\tau) &=& \overline{h}_{\lambda}(\vec{q},\tau) - 2 \ell_{P}^2 \int_{\tau_{i}}^{\tau} d \xi \, a^2(\xi) \,G[ q ( \xi- \tau)] \, 
\Pi_{\lambda}^{(X)}(\vec{q}, \xi), 
\nonumber\\
H_{\lambda}^{(X)}(\vec{q},\tau) &=& \overline{H}_{\lambda}(\vec{q},\tau) - 2 \ell_{P}^2  \int_{\tau_{i}}^{\tau} d \xi \, a^2(\xi) \,\widetilde{\,G\,}[ q ( \xi- \tau)] \, 
\Pi_{\lambda}^{(X)}(\vec{q}, \xi), 
\label{six12}
\end{eqnarray}
where $H^{(X)}_{\lambda} = \partial_{\tau} h^{(X)}_{\lambda}$ and $\overline{H}_{\lambda} =\partial_{\tau}\overline{h}_{\lambda} $; the overline distinguishes the (gauge-invariant) first-order contributions  from their second-order (gauge-dependent) counterparts.
After inserting Eq. (\ref{six7}) into Eqs. (\ref{six12}) the tensor amplitude $h_{\lambda}(\vec{q},\tau)$ follows by recalling the explicit expressions of the Green's functions  during the radiation-dominated stage i.e. 
\begin{equation}
G[q(\xi- \tau)] = -\frac{a(\xi)}{q \,a(\tau)} \sin{[q (\xi -\tau)]}, \qquad \widetilde{\,G\,}[q(\xi - \tau)] = \frac{a(\xi)}{a(\tau)} \cos{[q (\xi -\tau)]}.
\label{six13}
\end{equation}
To compute the effective energy density of the relic gravitons we now need to estimate first their 
energy density which ultimately depends on the form of the energy-momentum pseudo-tensor 
of the relic gravitons. For instance the energy-momentum pseudo-tensor obtained from the variation of the effective 
action of the relic gravitons with respect to the background metric leads to the energy density firstly derived by Ford and Parker
\cite{SIX,SEVEN}:
\begin{equation}
\rho_{gw} = \frac{1}{8 \ell_{\mathrm{P}}^2 a^2} \biggl[ \partial_{\tau} h_{k \ell}\, \partial_{\tau}h^{k \ell} + \partial_{m} h_{k\ell} \partial^{m} h^{k\ell}\biggr].
\label{six14}
\end{equation}
Recalling now Eqs. (\ref{onea6}), (\ref{six4}) and (\ref{six12}) the spectral energy density of the relic gravitons 
is obtained by taking the ratio between the average of Eq. (\ref{six14}) and the critical energy density 
according to a standard procedure\footnote{Mutatis mutandis this analysis coincides 
with the results of an analog problem involving the spectrum of gravitational radiation 
induced by waterfall fields \cite{EX2a} (see also \cite{EX3,EX4,EX5}).}; thus in our case the spectral energy density of the relic gravitons in critical units is given by:
\begin{eqnarray}
\Omega^{(X)}_{gw}(q,\tau) &=&  \frac{ q^2 \overline{P}_{T}(q)}{24\, H^2 \, a^2 \, |q\tau|^2} \biggl[ 1 + \frac{\sin{q\tau}}{q^2\tau^2} - \frac{\sin{2 q\tau}}{q\tau} \biggr]
+ \frac{q^3}{12}  \biggl(\frac{a_{1}^4 \, H_{1}^2}{a^4 \, H^2} \biggr)  \int_{-1}^{1} \, d\mu\, (1- \mu^2)^2
\nonumber\\
&\times&\int d k \, k^6\, \frac{\overline{P}_{\mathcal R}(k) \,\, \overline{P}_{\mathcal R}(|\vec{q} - \vec{k}|)}{k^3\,\,|\vec{q} - \vec{k}|^3}\, \biggl[ \overline{I}^{(X)\,2 }(\vec{k},\, \vec{q},\, \tau)+ \overline{J}^{(X)\,2 }(\vec{k},\, \vec{q},\, \tau) \biggr],
\label{six15}
\end{eqnarray}
where $\overline{I}^{(X)}(\vec{k},\, \vec{q},\, \tau)$ and $\overline{J}^{(X)}(\vec{k},\, \vec{q},\, \tau)$ are given by:
\begin{eqnarray}
\overline{I}^{(X)}(\vec{k},\, \vec{q},\, \tau) &=& \int_{\tau_{i}}^{\tau} \,\xi \, \sin{[q (\xi -\tau)]}\, M^{(X)}(k\, c_{st} \xi;\, |\vec{q} - \vec{k}|\, c_{st} \xi) \, d\xi,
\nonumber\\
\overline{J}^{(X)}(\vec{k},\, \vec{q},\, \tau) &=& \int_{\tau_{i}}^{\tau} \, \xi \, \cos{[q (\xi -\tau)]}\, M^{(X)}(k\, c_{st} \xi;\, |\vec{q} - \vec{k}|\, c_{st} \xi) \, d\xi.
\label{six16}
\end{eqnarray}
The spectral energy density of the relic gravitons  {\em inside the Hubble radius} in its full form 
(i.e. including the second-order corrections) follows from Eqs. (\ref{six15}) and (\ref{six16}) by recalling the limit of Eq. (\ref{six10}).  Thus the expressions of Eq. (\ref{six15}) (for $X = L$ and $X =U$) will eventually inherit a phase difference that however disappears after squaring and summing up the contributions of the two integrals (\ref{six16}) in each case. The common value of spectral energy density inside the 
sound horizon  is therefore 
\begin{eqnarray}
\Omega^{(U)}_{gw}(q,\tau_{0})= \Omega^{(L)}_{gw}(q,\tau_{0}) &=& \frac{r_{T} \, {\mathcal A}_{{\mathcal R}}\Omega_{R0} }{12}   \biggl(\frac{q}{q_{p}} \biggr)^{n_{T}}\biggl[ 1 + \frac{96 \, \pi^2 {\mathcal A}_{\mathcal R}}{5 r_{T}} f(n_{s}, q) \biggl(\frac{q}{q_{p}}\biggr)^{ 2 (n_{s} -1) - n_{T}} \biggr],
\label{six17}\\
f(n_{s}, q) &=& a_{1}(n_{s}) + a_{2}(n_{s})  \biggl(\frac{q_{p}}{q} \biggr)^{n_{s} +1} + 
a_{3}(n_{s}) \biggl(\frac{q_{max}}{q} \biggr)^{2n_{s} -5},
\label{six18}
\end{eqnarray}
where $\tau_{0}$ denotes the present value of the conformal time coordinate while $a_{i}(n_{s})$ (with 
$i=1,\,2\,3$) are three numerical constants\footnote{Even if the explicit expressions are immaterial for the 
present discussion we have that $a_{1}(n_{s}) = (n_{s} -6)/[(2 n_{s} -5) (n_{s}+1)]$, $ a_{2}(n_{s}) =  - 1/(n_{s} +1)$
and  $a_{3}(n_{s}) = 1/(2 n_{s} -5)$.}. The expressions of the coefficients $a_{i}(n_{s})$ follow from the integration of Eq. (\ref{six15}) 
first over $\mu$ and then over $k$ between $q_{p}$ and $q_{max}$. The integration over $k$ can be approximated in two 
separate regions (i.e. $k<\,q$ and $k>\,q$); this way of approximating the integrals compares quite well with the 
numerical results as explicitly discussed in the case of waterfall fields where the power spectra appearing in the convolutions  have larger slopes but similar analytical expressions. Since $\nu_{p} = 2 \pi q_{p}$ is in the aHz region (see discussion after Eq. (\ref{six3})) 
 and $\nu_{max} = 2\pi q_{max} =190$ MHz we have that $f(n_s,\,q)= {\mathcal O}(10^{-2})$  for typical scalar 
 spectral indices $0.9< n_{s}< 1$ .

Let us finally consider Eqs. (\ref{six15}) and (\ref{six16}) when the corresponding wavelengths are outside the sound horizon. 
Once again, with the help of these asymptotic expressions 
the integrals $\overline{I}^{(X)}(\vec{k},\, \vec{q},\, \tau)$ and $\overline{J}^{(X)}(\vec{k},\, \vec{q},\, \tau)$ of Eq. (\ref{six16}) can be estimated. The first-order contribution has the standard form valid during the radiation-dominated phase and it follows from the first term at the right-hand side of Eq. (\ref{six15}) for $q\tau \ll 1$;
the second-order correction is however different in the two gauges so that the general form of $\Omega_{gw}^{(X)}(q,\tau)$ is:
 \begin{equation}
\Omega^{(X)}_{gw}(q,\tau) =\overline{\Omega}_{gw}(q,\tau)\biggl[ 1 + \omega^{(X)}_{gw}(q, \tau) \biggr],\qquad \overline{\Omega}_{gw}(q,\tau) =  \frac{r_{T}{\mathcal A}_{{\mathcal R}}}{12} q^2 \tau^2 \biggl(\frac{q}{q_{p}}\biggr)^{n_{T}},
 \label{six19}
 \end{equation}
 where the two functions $\omega^{(L)}(q,\tau_{0})$ and $\omega^{(U)}(q,\tau_{0})$ are: 
 \begin{eqnarray}
\omega^{(L)}_{gw}(q,\tau) &=& \frac{64}{15} \frac{{\mathcal A}_{\mathcal R}}{r_{T}} \, \Omega_{R0} \, q^2 \tau^2 \, \biggl[ 1 + \frac{q^2 \tau^2}{9} \biggr]\, \biggl(\frac{q}{q_{p}}\biggr)^{2(n_{s} -1) -n_{T}} \overline{f}^{(L)}(n_{s},\, q),
\nonumber\\
 \omega^{(U)}_{gw}(q,\tau) &=& \frac{4}{135} \frac{{\mathcal A}_{\mathcal R}}{r_{T}} \, \Omega_{R0} \, q^6 \tau^6 \, \biggl[ 1 + \frac{q^2 \tau^2}{25} \biggr]\, \biggl(\frac{q}{q_{p}}\biggr)^{2(n_{s} -1)-n_{T}} \overline{f}^{(U)}(n_{s},\, q). 
 \label{six20}
\end{eqnarray}
The form of  $\overline{f}^{(L)}(n_{s},\, q)$ and $\overline{f}^{(U)}(n_{s},\, q)$ is not central to the present discussion and it is anyway similar to $f(n_{s}, q)$ appearing in Eq. (\ref{six18}). What matters here
is the parametric dependence of the correction upon $q\,\tau$, i.e. $ \omega^{(U)}_{gw}(q,\tau)/\omega_{gw}^{(L)}(q,\tau) =  {\mathcal O}(|q\tau|^4)$. 

All in all we have that the results obtained in the case of  are fully consistent with the ones obtained in Eqs. (\ref{five7}) and (\ref{five20}). In particular Eq. (\ref{six17}) corresponds to Eq. (\ref{five7}) and demonstrates that the spectral energy 
densities computed in different gauges coincide when the wavelengths of the scalar modes are inside the sound horizon. 
Equation (\ref{five20}) corresponds instead to Eq. (\ref{six20}) with the caveat that Eq. (\ref{five20}) applies for the 
anisotropic stresses while the spectral energy density of Eq. (\ref{six20}) is instead quadratic in the effective anisotropic stress. This is why the mismatch between the two expressions 
is not given by $|q\tau|^2$ (as in Eq. (\ref{five20})) but by the square of it (i.e. $|q\tau|^4$). 
We conclude that the limit of the general expressions obtained without specifying the background geometry 
coincide, as expected, in the particular case of a radiation-dominated plasma once the corresponding 
expressions are evaluated either inside or outside the sound horizon.

\renewcommand{\theequation}{7.\arabic{equation}}
\setcounter{equation}{0}
\section{Effective anisotropic stress from the second-order action}
\label{sec7}
So far  we suggested a gauge-invariant method to compare gauge-dependent 
results. The idea is to choose a coordinate system where the gauge freedom is completely 
fixed and to compute the effective anisotropic stress in that particular gauge. At the very end 
the results will then be expressed in terms of the gravitating normal modes of the 
system and of their conformal time derivatives. Since the gravitating normal modes 
obey {\em the same} evolution equations in different coordinate system, the results 
obtained in various gauges are most easily assessed. In general terms 
the gravitating normal modes will depend on the total anisotropic stress and on 
the non-adiabatic pressure fluctuations. The systematic use of the WKB approximation 
demonstrated that when the wavelengths are shorter than the sound horizon the results 
of different gauges are all consistent while in the opposite regime they are not.
A similar problem arose in the past when discussing the effective energy density 
and pressure of the gravitational field: different strategies lead in fact to 
consistent results only inside the Hubble radius but not outside of it.  In what 
follows we intend to suggest that probably the best way of defining the effective anisotropic stress 
is to start from the second-order action of the curvature inhomogeneities in the same way as the 
simplest way of defining the energy density of the relic gravitons is to start from their 
second-order action.

\subsection{The effective energy density of the relic gravitons}
Let us start by briefly examining the derivation of the energy-momentum pseudo-tensor of the 
relic gravitons \cite{SIX,SEVEN}. Since the effective action of the relic gravitons is:
\begin{equation}
S_{t} = \frac{1}{8 \ell_{P}^2} \int d^{4} x \sqrt{ - \overline{g}} \,\, \overline{g}^{\alpha\beta} \, \, \partial_{\alpha} h_{ij} \, \partial_{\beta} h^{ij},
\label{seven1}
\end{equation}
the associated energy-momentum pseudo-tensor can be introduced from the functional derivative 
of $S_{t}$ with respect to $\overline{g}_{\mu\nu}$ by considering $h_{ij}$ and $\overline{g}_{\mu\nu}$ as 
independent variables \cite{SIX,SEVEN}. From Eq. (\ref{seven1}) the explicit form of the energy-momentum pseudo-tensor 
is
 \begin{equation}
T_{\mu\nu} = \frac{1}{4 \ell_{\mathrm{P}}^2} \biggl[ \partial_{\mu} h_{i j} \,\,\partial_{\nu} h^{i j} 
- \frac{1}{2} \overline{g}_{\mu \nu} \,\,\biggl(\overline{g}^{\alpha\beta}\, \partial_{\alpha} h_{ij} \,\,\partial_{\beta} h^{ij} \biggr)\biggr],
\label{seven2}
\end{equation}
and it can be derived by computing the variation of $S_{t}$ with respect to $\delta \overline{g}^{\mu\nu}$ i.e.
\begin{equation}
\delta S_{t} = \frac{1}{2} \int d^{4} x \,\sqrt{- \overline{g}}\, T_{\mu\nu}\, \delta \overline{g}^{\mu\nu}, \qquad T_{\mu}^{\nu} = \overline{g}^{\alpha\nu} T_{\alpha\nu}.
\label{seven2a}
\end{equation}
Since the indices of $T_{\mu\nu}$  are raised and lowered with the help of the background metric, the energy density and the pressure are defined from the various components of the energy-momentum pseudo-tensor as:
\begin{eqnarray}
T_{0}^{0} &=& \rho_{gw}, \qquad T_{i}^{0} = S_{i}^{(F)}= 
\frac{1}{4 \ell_{\mathrm{P}}^2 a^2} \partial_{\tau} h_{k\ell} \,\partial_{i} h^{k\ell},
\nonumber\\
T_{i}^{j} &=& - p_{gw} \,\,\delta_{i}^{j} + \Pi_{gw\, i}^{\,\,j},
\label{seven3}
\end{eqnarray}
where  $\rho_{gw}$ and $p_{gw}$ are the energy density and the pressure of the relic gravitons:
\begin{eqnarray}
\rho_{gw} &=& \frac{1}{8 \ell_{\mathrm{P}}^2 a^2} \biggl[ \partial_{\tau} h_{k \ell}\, \partial_{\tau}h^{k \ell} + \partial_{m} h_{k\ell} \partial^{m} h^{k\ell}\biggr],
\label{seven4}\\
p_{gw} &=&   \frac{1}{8 \ell_{\mathrm{P}}^2 a^2} \biggl[ \partial_{\tau} h_{k\ell}\partial_{\tau} h^{\,k\ell} - \frac{1}{3} \partial_{m} h_{k \ell} \,\partial^{m} h^{\,k\ell}  \biggr].
\label{seven5}
\end{eqnarray}
The energy density obtained in Eq. (\ref{seven4})  coincides in fact with the result already mentioned in Eq. (\ref{six14}).
In Eq. (\ref{seven3}) we have a further class of traceless anisotropic stresses (i.e. $\Pi_{i}^{\,\,i} =0$), namely the anisotropic stress 
of the tensor modes:
\begin{equation}
\Pi_{gw\,i}^{\,\,j} = \frac{1}{4 \ell_{\mathrm{P}}^2 a^2} \biggl[ - \partial_{i} \,h_{k\ell} \,\,\partial^{j} \,h^{k\ell} + \frac{1}{3} \delta_{i}^{j} \,\,\partial_{m}\, h_{k \ell} \,\,\partial_{m} \,h_{k\ell} \biggr].
\label{seven6}
\end{equation}
Equation (\ref{seven6}) accounts for the anisotropic stress induced by the tensor modes. What we are looking for is 
the anisotropic stress induced by the scalar modes of the geometry.

\subsection{The effective anisotropic stress in the case of an irrotational fluid}
The evolution of the gravitating normal modes of Eq. (\ref{one1}) can be derived from an action that is very similar 
to the one of Eq. (\ref{seven1}):
\begin{equation}
S_{{\mathcal R}} = \frac{1}{2} \int d^{3} x \int d\tau \, z_{t}^2 \biggl[ \partial_{\tau} {\mathcal R} \partial_{\tau} {\mathcal R} - c_{st}^2 \, \partial_{k} {\mathcal R} \, \partial^{k} {\mathcal R} \biggr]. 
\label{seven7}
\end{equation}
Recalling the explicit form of $z_{t}$ (i.e. Eq. (\ref{one2})) and using the background equations(\ref{one2a}) the action ${\mathcal S}_{{\mathcal R}}$ can also be expressed as: 
\begin{equation}
S_{{\mathcal R}} = \frac{1}{\ell_{P}^2} \int d^{3} x \int d\tau \, a^2 \, \biggl(\frac{{\mathcal H}^2 - {\mathcal H}^{\prime}}{{\mathcal H}^2}\biggr) \biggl[ \frac{1}{c_{st}^2}\partial_{\tau} {\mathcal R} \partial_{\tau} {\mathcal R} -   \partial_{k} {\mathcal R} \, \partial^{k} {\mathcal R} \biggr],
\label{seven8}
\end{equation}
so that the effective anisotropic stress now becomes
\begin{equation}
\Pi_{i}{\,\,\,j} = - \frac{2 ({\mathcal H}^2 - {\mathcal H}^{\prime})}{a^2 {\mathcal H}^2 \ell_{P}^2} \biggl[ \partial_{i} \,{\mathcal R} \partial^{j} {\mathcal R} - 
\frac{1}{3}\partial_{k} {\mathcal R}\, \partial^{k} {\mathcal R}\, \delta_{i}^{\,\,j}\biggr]. 
\label{seven9}
\end{equation}
It is quite clear that the effective anisotropic stress of Eq. (\ref{seven9}) gives exactly the leading-order 
contribution already deduced in the $L$-gauge and in the $U$-gauge.  In particular if we Fourier transform Eq. (\ref{seven9}) 
and project it along the tensor polarization we will have that\footnote{Note that the term proportional to $\delta_{ij}$ 
appearing in Eq. (\ref{seven9}) does not contributed to $\Pi_{\lambda}$.} 
\begin{eqnarray}
\Pi_{\lambda}(\vec{q}, \tau) = - \frac{ ({\mathcal H}^2 - {\mathcal H}^{\prime})}{(2\pi)^{3/2}\,\ell_{P}^2\, a^2(\tau)\,{\mathcal H}^2 } \int\, d^{3}k\, \,k^2\,\, s_{\lambda}(\hat{k}, \hat{q})\, {\mathcal R}_{\vec{k}} \, {\mathcal R}_{\vec{q} - \vec{k}}
\label{seven10}
\end{eqnarray}
where $s_{\lambda}(\hat{k},\hat{q}) = \hat{k}_{i} \, \hat{k}_{j} \, e^{i \, j}_{\lambda}(\hat{q})$. While Eq. (\ref{seven10}) 
coincides with the leading-order expression obtainable in specific gauges inside the sound horizon, outside of it 
this expression is the same. We therefore suggest that the second-order action of the scalar modes 
could be directly used to deduce the effective anisotropic stress of the relic gravitons.

\subsection{The effective anisotropic stress in the case of scalar field matter}
To corroborate even further the conclusions of the previous paragraph let us consider the case of 
scalar field matter. In this case the curvature perturbations obey the following effective action 
\begin{equation}
S_{{\mathcal R}} = \frac{1}{2} \int d^{4} x \biggl(\frac{\varphi^{\prime}}{{\mathcal H}}\biggr)^2 \, \sqrt{- \overline{g}} \, \overline{g}^{\alpha\beta} \partial_{\alpha} {\mathcal R} \, \partial_{\beta} {\mathcal R}.
\label{seven11}
\end{equation} 
The background equation $\varphi^{\prime\, 2} = 2 ({\mathcal H}^2 - {\mathcal H}^{\prime})/\ell_{P}^2$ can be used in Eq. (\ref{seven11}) and the action becomes 
\begin{equation}
S_{{\mathcal R}} = \frac{1}{\ell_{P}^2} \int d^{4} x \biggl(\frac{{\mathcal H}^2 - {\mathcal H}^{\prime}}{{\mathcal H}^2}\biggr)\, \sqrt{- \overline{g}} \, \overline{g}^{\alpha\beta} \partial_{\alpha} {\mathcal R} \, \partial_{\beta} {\mathcal R}.
\label{seven12}
\end{equation} 
 By taking the functional derivative with respect to $\overline{g}^{\mu\nu}$ we have that the energy-momentum pseudo-tensor of the curvature inhomogeneities is:
\begin{eqnarray}
{\mathcal T}_{0}^{0} = \rho_{{\mathcal R}},\qquad 
{\mathcal T}_{i}^{\,\,\,j} = - p_{{\mathcal R}}\,\,\delta_{i}^{\,\,\,j}  + \Pi_{i}^{\,\,\,\,j},
\label{seven13}
\end{eqnarray}
where $\rho_{{\mathcal R}}$, $p_{{\mathcal R}}$ and $ \Pi_{i}^{\,\,\,\,j}$ are given, respectively, by:
\begin{eqnarray}
\rho_{{\mathcal R}} &=& \frac{\varphi^{\,\prime \,\,2}}{2 \,{\mathcal H^2} \, a^2}
\biggl[\partial_{\tau} {\mathcal R} \partial_{\tau} {\mathcal R} + \partial_{k} {\mathcal R} \partial^{k} {\mathcal R} \biggr], 
\nonumber\\
p_{{\mathcal R}} &=& \frac{\varphi^{\,\prime \,\,2}}{2 \,{\mathcal H^2} \, a^2}
\biggl[\partial_{\tau} {\mathcal R} \partial_{\tau} {\mathcal R} - \frac{1}{3} \partial_{k} {\mathcal R} \partial^{k} {\mathcal R} \biggr],
\label{seven14}\\
 \Pi_{i}^{\,\,\,\,j} &=& - \frac{2( {\mathcal H}^2 - {\mathcal H}^{\prime})}{\, \ell_{P}^2\, {\mathcal H^2} \, a^2} \biggl[\partial_{i} {\mathcal R}\, \partial^{j} {\mathcal R} - \frac{\delta_{i}^{\,\,\,j}}{3} \partial_{k} {\mathcal R} \partial^{k} {\mathcal R} \biggr].
\label{seven15}
\end{eqnarray} 
By projecting Eq. (\ref{seven15}) over the tensor polarizations we obtain the same result of Eq. (\ref{seven10}).

\newpage

\renewcommand{\theequation}{.\arabic{equation}}
\setcounter{equation}{0}
\section{Concluding remarks and general lessons}
\label{sec8}
The starting point of this analysis has been the observation that the effective anisotropic stresses 
induced by the scalar modes of the geometry depends on the coordinate system where it is 
evaluated. Not all the coordinate systems are equally viable: the ones where the gauge freedom is completely 
eliminated guarantee the absence of spurious gauge modes and this is why the attention 
has been focussed on the longitudinal and on the uniform curvature gauges. In spite of this 
important difference the anisotropic stresses computed in different coordinate systems 
depend on the evolution of the pivotal variables of that particular gauge. 

To avoid this drawback we suggested how the gauge-dependent results 
could be compared in a gauge-invariant manner. By this we simply stress that 
the results obtained in diverse coordinate systems  can only be compared in a meaningful way 
by expressing the gauge-dependent results in terms of the gravitating 
normal modes of the system. This is the novel idea proposed and scrutinized in this paper. 
Since the gravitating normal modes of the plasma obey the same evolution equation in any coordinate system 
there will be a unique evolution equation determining the effective anisotropic stresses in different gauges. 
The results of this analysis are, in short, the following:
\begin{itemize}
\item{} the evolution of the gauge-invariant curvature inhomogeneities has been analyzed 
in general terms by including the non-adiabatic pressure fluctuations and the scalar 
anisotropic stress;
\item{} inside the sound horizon the effective anisotropic stresses computed in the $L$-gauge and in the $U$-gauge 
coincide to leading order (i.e. they are gauge-invariant from the practical viewpoint);
\item{} for typical wavelengths larger than the sound horizon the evolution of the normal modes 
imply instead that the anisotropic stresses are sharply different and that, in particular, the result 
in the $U$-gauge is much smaller than the one in the $L$-gauge;
\item{} even if the present approach employs the WKB approximation (and does not assume any specific background evolution) 
the obtained results have been explicitly corroborated by  the analysis of a radiation 
dominated plasma;
\item{} we finally argued that the effective anisotropic stress of the curvature inhomogeneities can be obtained from the functional derivative of the second-order action of curvature inhomogeneities with respect to the background metric. 
\end{itemize}
The obtained results suggest therefore that the effective anisotropic stresses 
are approximately gauge-invariant inside the sound horizon but sharply different outside of it.
 The same kind of spurious gauge-invariance examined here
is also manifest when the energy density of the relic gravitons is derived from competing 
energy-momentum pseudo-tensors. To lowest order the ambiguity can 
be solved (or alleviated) by selecting an energy-momentum pseudo-tensor with reasonable 
physical properties such as the one obtained long ago by Ford and Parker. The present considerations 
show however that some ambiguities are likely to reappear from the higher-order processes as 
a direct consequence of the lack of localization of the energy-momentum of the gravitational field. 
Following the same logic that leads to the energy-momentum pseudo-tensor of the relic gravitons,
we can use the second-order action of the scalar modes to obtain the effective anisotropic stress.
It turns out that the results obtained in this way coincide (inside 
the sound horizon) with the expressions derived in different coordinate systems where the gauge freedom is completely fixed. When the wavelengths of the curvature inhomogeneities are larger than the sound horizon the gauge-dependent results are sharply different;  
the second-order action leads instead to an expression that formally coincide with 
the result valid inside the sound horizon. 

\section*{Acknowledgements}
The author wishes to thank T. Basaglia, A. Gentil-Beccot, S. Rohr and J. Vigen of the CERN 
Scientific Information Service for their kind assistance.


\newpage 


\begin{thebibliography}{99}

\itemsep -2pt

\bibitem{ONE} S. Weinberg, {\it Gravitation and Cosmology}, (Wiley, New York, 1972).

\bibitem{TWO} L. D. Landau and E. M. Lifshitz, {\it The Classical Theory of Fields}, (Pergamon Press, New York, 1971).

\bibitem{TWOa} M.~Giovannini, Prog. Part. Nucl. Phys. {\bf 112}, 103774 (2020).

 \bibitem{THREE} D.~R.~Brill and J.~B.~Hartle, Phys.\ Rev.\  {\bf 135}, B271 (1964).

\bibitem{FOUR}  R.~A.~Isaacson,  Phys.\ Rev.\  {\bf 166}, 1263 (1968);
 Phys.\ Rev.\  {\bf 166}, 1272 (1968).
 
 \bibitem{FIVE} M. A. H. MacCallum and A. H. Taub, Commun. Math. Phys. {\bf 30}, 153 (1973).

\bibitem{SIX} L. H. Ford and L. Parker, Phys. Rev. {\bf D16}, 1601 (1977); Phys.\ Rev.\ D {\bf 16}, 245 (1977).

\bibitem{SEVEN} M.~Giovannini,  Phys.\ Rev.\ D {\bf 99}, 083501 (2019).

\bibitem{EIGHT} L. R. Abramo, Phys Rev. D {\bf 60}, 064004 (1999).

\bibitem{EIGHTA}  M. Giovannini, Phys.\ Rev.\ D {\bf 73} 083505 (2006).

\bibitem{EIGHTB} D. Su and Y. Zhang, Phys.\ Rev.\ D {\bf 85}, 104012 (2012).

 \bibitem{EIGHTC}  L.~C.~Stein and N.~Yunes,  Phys.\ Rev.\ D {\bf 83}, 064038 (2011); 
 M.~Isi and L.~C.~Stein,  Phys.\ Rev.\ D {\bf 98}, 104025 (2018).

\bibitem{NINE} S.~V.~Babak and L.~P.~Grishchuk, Phys.\ Rev.\ D {\bf 61}, 024038 (2000).

\bibitem{TEN} L.~M.~Butcher, A.~Lasenby and M.~Hobson,  Phys.\ Rev.\ D {\bf 78}, 064034 (2008); 
Phys.\ Rev.\ D {\bf 80}, 084014 (2009).

\bibitem{ELEVEN}  L.~M.~Butcher, M.~Hobson and A.~Lasenby,
  Phys.\ Rev.\ D {\bf 82}, 104040 (2010); Phys.\ Rev.\ D {\bf 86}, 084012 (2012)

\bibitem{TWELVE} M.~Giovannini,  Phys.\ Rev.\ D {\bf 91},  023521 (2015).  
  
\bibitem{THIRTEEN} K.~N.~Ananda, C.~Clarkson and D.~Wands, Phys.\ Rev.\ D {\bf 75}, 123518 (2007).

\bibitem{SIXTEEN} J.~Hwang, D.~Jeong and H.~Noh, Astrophys.\ J.\  {\bf 842},  46 (2017).

\bibitem{SEVENTEEN} R. g. Cai, S. Pi and M. Sasaki, Phys. Rev. Lett. 122, 201101 (2019).

\bibitem{EIGHTEEN}  C.~Yuan, Z.~Chen and Q.~Huang, Phys.\ Rev.\ D {\bf 101},  063018 (2020).

\bibitem{NINETEEN} K.~Inomata and T.~Terada, Phys.\ Rev.\ D {\bf 101},  023523 (2020).

\bibitem{TWENTY} K.~Tomikawa and T.~Kobayashi, Phys. Rev. D {\bf 101},  083529 (2020).

\bibitem{TWENTYONEa} M.~Giovannini, [arXiv:2005.04962 [hep-th]].

\bibitem{TWENTYONEb}  M.~Giovannini, [arXiv:2006.02760 [gr-qc]].

\bibitem{TWENTYONEc} C.~P.~Ma and E.~Bertschinger,  Astrophys.\ J.\  {\bf 455}, 7 (1995).

\bibitem{TWENTYFOURa} S. Weinberg, Phys. Rev. D {\bf 69},  023503 (2004).

\bibitem{TWENTYFOURb}  B. A. Stefanek and W. W. Repko,  Phys. Rev. D {\bf 88}, 083536 (2013).

\bibitem{TWENTYTWO} V.~N.~Lukash,  Sov.\ Phys.\ JETP {\bf 52}, 807 (1980) [Zh. Eksp. Teor. Fiz. {\bf 79}, 1601 (1980)].

\bibitem{TWENTYTWOa} E.~M.~Lifshitz Zh. Eksp. Teor. Fiz. {\bf 16}, 587 (1946) [J. Phys. (USSR) {\bf 10}, 116 (1946)].

\bibitem{TWENTYTWOb} E.~M.~Lifshitz and I.~M.~Khalatnikov, Adv.\ Phys.\  {\bf 12}, 185 (1963). 

\bibitem{TWENTYTWOc} J. Bardeen, Phys. Rev. {\bf D22}, 1882 (1980).

\bibitem{TWENTYTWOd}  H.~Kodama, M.~Sasaki,  Prog.\ Theor.\ Phys.\ Suppl.\  {\bf 78}, 1 (1984).

\bibitem{TWENTYTWOe} G. Chibisov, V. Mukhanov,  Mon.\ Not.\ Roy.\ Astron.\ Soc.\  {\bf 200}, 535 (1982); V. Mukhanov,  Sov.\ Phys.\ JETP {\bf 67}, 1297 (1988)  [Zh. Eksp. Teor. Fiz. {\bf 94}, 1 (1988)].

\bibitem{TWENTYTWOf} J. Bardeen, P. Steinhardt, and M. Turner, Phys. Rev. {\bf D28}, 679 (1983); 
J.~A.~Frieman and M.~S.~Turner, Phys.\ Rev.\ D {\bf 30}, 265 (1984).

\bibitem{TWENTYTWOg} K.~Enqvist, H.~Kurki-Suonio and J.~Valiviita, Phys.\ Rev.\  D {\bf 62}, 103003 (2000);  
J.~Valiviita and V.~Muhonen, Phys.\ Rev.\ Lett.\  {\bf 91}, 131302 (2003).

\bibitem{TWENTYTWOh}  H.~Kurki-Suonio, V.~Muhonen and J.~Valiviita, Phys.\ Rev.\  D {\bf 71}, 063005 (2005); 
R.~Keskitalo, H.~Kurki-Suonio, V.~Muhonen and J.~Valiviita,  JCAP {\bf 0709}, 008 (2007).

\bibitem{TWENTYTWOi} M.~Giovannini,  Phys.\ Rev.\ D {\bf 74}, 063002 (2006);  Class.\ Quant.\ Grav.\  {\bf 23}, 4991 (2006);
  PMC Phys.\ A {\bf 1}, 5 (2007); M.~Giovannini and K.~E.~Kunze,  Phys.\ Rev.\ D {\bf 77}, 063003 (2008); 

\bibitem{TWENTYFIVEa} W. Press and E. Vishniac Astrophys. J. {\bf 239} 1 (1980); 
Astrophys. J. {\bf 236}, 323 (1980).

\bibitem{TWENTYTHREE}  M. Giovannini, Phys. Rev. D  {\bf 87}, 083004 (2013).

\bibitem{EX1} M. Abramowitz and I. A. Stegun, {\it Handbook of Mathematical Functions} (Dover, New York, 1972).

\bibitem{EX2}  I. S. Gradshteyn and I. M. Ryzhik,  
{\it Tables of Integrals, Series and Products (fifth edition)},  (Academic Press, New York, 1994).

\bibitem{EX2a}  M.~Giovannini,  Phys.\ Rev.\ D {\bf 82}, 083523 (2010).

\bibitem{EX3} J.~Fonseca, M.~Sasaki and D.~Wands,  JCAP {\bf 1009}, 012 (2010).

\bibitem{EX4} D.~H.~Lyth,  Prog.\ Theor.\ Phys.\ Suppl.\  {\bf 190}, 107 (2011).

\bibitem{EX5} S.~Clesse,  Phys.\ Rev.\ D {\bf 83}, 063518 (2011).



\end{thebibliography}
\end{document}